\documentclass[pdflatex,sn-aps]{sn-jnl}

\usepackage{graphicx}%
\usepackage{multirow}%
\usepackage{amsmath,amssymb,amsfonts}%
\usepackage{amsthm}%
\usepackage{mathrsfs}%
\usepackage[title]{appendix}%
\usepackage{xcolor}%
\usepackage{textcomp}%
\usepackage{manyfoot}%
\usepackage{booktabs}%
\usepackage{listings}%

\usepackage{setspace}
\usepackage{lipsum}
\begin{document}

\title[Improving relativistic energy density functionals with tensor couplings]{Improving relativistic energy density functionals with tensor couplings}


\author*[1,2,3]{\fnm{Stefan} \sur{Typel}}\email{stypel@ikp.tu-darmstadt.de,\\
  ORCID: 0000-0003-3238-9973}

\author[4]{\fnm{Shalom} \sur{Shlomo}}\email{s-shlomo@tamu.edu,\\
  ORCID: 0000-0002-4049-7563}


\affil*[1]{\orgname{Technische Universit\"{a}t Darmstadt, Fachbereich Physik}, \orgdiv{Institut f\"{u}r Kernphysik}, \orgaddress{\street{Schlossgartenstra\ss{}e 9}, \city{Darmstadt}, \postcode{64289}, 
    \country{Germany}}}

\affil[2]{\orgname{GSI Helmholtzzentrum f\"{u}r Schwerionenforschung}, \orgdiv{Theorie},  \orgaddress{\street{Planckstra\ss{}e 1}, \city{Darmstadt}, \postcode{64291}, 
    \country{Germany}}}

\affil[3]{\orgdiv{Helmholtz Forschungsakademie Hessen f\"{u}r FAIR (HFHF)}, \orgname{GSI Helmholtzzentrum f\"{u}r Schwerionenforschung, Campus Darmstadt},  \orgaddress{\street{Schlossgartenstra\ss{}e 2}, \city{Darmstadt}, \postcode{64289}, 
    \country{Germany}}}

\affil[4]{\orgname{Texas A\&{}M University}, \orgdiv{Cyclotron Institute},
  \orgaddress{
  \city{College Station}, \postcode{77843}, \state{Texas}, \country{USA}}}


\abstract{Energy density functionals (EDFs)
    have been used extensively
    with great success to calculate properties
    of nuclei and to predict the equation of state (EOS)
    of dense nuclear matter.
    Besides non-relativistic EDFs, mostly of the Skyrme or Gogny type,
    relativistic EDFs of different types are in widespread use. In these
    latter approaches, the effective in-medium interaction is described by an
    exchange of mesons between nucleons. In most cases, only minimal
    meson-nucleon couplings are considered. The effects of additional
    tensor couplings were rarely investigated. In this work, a new relativistic
    EDF with tensor couplings and density dependent minimal meson-nucleon
    couplings will be presented. The parameters of the model are
    determined using a carefully selected set of experimental data with
    realistic uncertainties that are determined self-consistently. Predictions
    for various nuclear observables, the nuclear matter
    equation of state, and properties of neutron stars
    are discussed.}

\keywords{Relativistic energy density functional, tensor couplings, parameterisation, nuclear properties, nuclear matter, equation of state, neutron stars}



\maketitle

\section{Introduction}

For more than half a century, energy density functionals (EDFs) have been used with great success in the description of atomic nuclei across the nuclear chart and nuclear matter over a large range of densities and isospin asymmetries. Many improvements were introduced in these phenomenological models leading to a more and more precise prediction of nuclear observables but, nevertheless, the quest for further improvements remains. There are different strategies to develop more accurate EDFs. They either follow a systematic approach, e.g., using expansions in power of small quantities, or try to incorporate known effects in a more heuristic way.

EDFs exist in different versions. Most well known are non-relativistic approaches that employ convenient representations of effective in-medium interactions between nucleons, e.g., the Skyrme or Gogny type models
\cite{Agrawal:2005ix,Stone:2006fn,Dutra:2012mb,Sellahewa:2014nia}.
Relativistic or covariant EDFs start from a field-theoretic approach where the interaction is described by meson-exchange taking medium dependent effects into account in different ways
\cite{Reinhard:1986qq,Rufa:1988zz,Reinhard:1989zi,Serot:1992ti,Fuchs:1995as,Ring:1996qi,Typel:1999yq,Vretenar:2005zz,Meng:2005jv,Niksic:2011sg,Dutra:2014qga}.
Either self-interactions of the mesons are introduced in so-called nonlinear (NL) models or a density dependence (DD) of the the meson-nucleon couplings is utilized.
Another class of relativistic EDFs uses only nucleons as degrees
of freedom with point couplings and strengths
that are constants or depend on the density, see, e.g., Refs.\
\cite{Nikolaus:1992zz,Burvenich:2001rh,Niksic:2008vp,Zhao:2010hi}.
No explicit meson contributions appear in these formulations.
These extensions of the original Walecka model \cite{Walecka:1974qa}
are essential to obtain reasonable properties of nuclei and nuclear matter.
Applying certain approximations to the full many-body wave function of the system of interest, e.g., of Hartree, Hartree-Fock or Hartree-Fock-Bogoliubov type, a set of coupled equations is found which has
to be solved self-consistently.

In the present work, new parameterisations of a relativistic EDF
will be presented that include tensor couplings between nucleons
and mesons in addition to the standard minimal couplings.
Such an extension has been considered before in early applications of relativistic EDFs, see, e.g., Ref.\ \cite{Rufa:1988zz},
but it has been explored scarcely in depth.
Effects of considering $\omega$-tensor couplings in EDFs
on the spin-orbit splittings and indirectly on
the nucleon effective mass have been studied
in Ref.\  \cite{Furnstahl:1997tk}.
A relativistic EDF with $\omega$ and $\rho$-tensor couplings in the
framework of the NL coupling models has been developed in Ref.\
\cite{Mao:2002xm} taking into account also contributions from Dirac-sea
nucleons. The parameters of this model were adjusted to the energies
of a eight nuclei from ${}^{16}$O to ${}^{208}$Pb.
Including $\rho$-tensor, but no $\omega$-tensor,
contributions in relativistic Hartree-Fock calculations showed that
the single-nucleon spectrum could be improved \cite{Long:2007dw}.
Tensor couplings were also included
in the framework of point couplings models
in the Hartree-Fock approximation, see, e.g., Ref.\ \cite{Zhao:2022xhq}.
The interest in relativistic EDFs with tensor couplings has been grown recently as it can affect the description of surface properties of nuclei and (indirectly) the equation of state of nuclear matter.
New parametrisations were developed in Ref.\
\cite{Typel:2020ozc} for relativistic models with meson-nucleon couplings
that depend on vector or scalar densities of the medium in various combinations,
but using a more limited number of nuclear observables in the parameter fit
as compared to the present work.
A new parameterisation of a covariant EDF with density
dependent minimal couplings and  density independent
tensor couplings in the Hartree-Bogoliubov approximation
was developed recently \cite{Mercier:2022svw}. The model parameters
were determined using constraints from nuclear matter and nuclei with
preset uncertainties of the fitted observables.
The effects of tensor couplings on different properties 
of nuclei and, using the resulting equation of state (EOS) of dense matter, on neutron stars were explored in Ref.\ \cite{Salinas:2023qic}, however without a thorough determination of model parameters. Here, we shall embark on the development of a new relativistic EDF with minimal and tensor couplings using a careful selection of nuclear observables that are used in the fit of the model parameters.
Particular attention is paid to the choice
of the uncertainties of the considered observables.
The medium dependence of the interaction is taken into account with a density dependence of the minimal couplings using the functional forms introduced in
Ref.\ \cite{Typel:1999yq}.
In addition to the description of nuclei, the implications
of the new parameterisations on the equation of state of nuclear matter
and the properties of neutron stars will be explored.

The content of this paper is structured as follows:
The formalism of the relativistic energy density functional is presented
in Section \ref{sec:REDF}. The main quantities are introduced and 
the notation is established, in particular regarding the contribution
with the tensor couplings. This section also provides the relevant equations
for the description of spherical nuclei and nuclear matter in two subsections.
In Section \ref{sec:para} the approach to determine the model parameters
is outlined. The functional form of the density dependent couplings is
specified and the selection of the observables and nuclei in the fit is
presented. Details of the fitting strategy are also given. Results and their
discussion follow in Section \ref{sec:res}, where properties of nuclei,
nuclear matter, and neutron stars are considered
in individual subsections. Finally, we close with a summary and outlook
in Section \ref{sec:concl}.
  
\section{Relativistic energy density functional}
\label{sec:REDF}

In a relativistic theory of nuclei and nuclear matter, nucleons and mesons are assumed to be the relevant degrees of freedom. Protons and neutrons are represented as Dirac spinors $\Psi_{\eta}$ (with $\eta=\pi$ for protons and $\eta=\nu$ for neutrons) following Fermi statistics, whereas mesons are described by bosonic fields. The strong interaction between nucleons can be represented effectively by an exchange of mesons of different type. They often share their properties with those mesons of the same name and quantum numbers that are observed in experiments but are not necessarily identical. Actually, it is possible to describe the main features of the strong interaction by a rather small set of effective mesons that reduces the complexity of the calculations and the number of model parameters.

In relativistic EDFs, it is common practise to introduce an iso-scalar, Lorentz-scalar $\sigma$ meson to model the attraction and an iso-scalar, Lorentz-vector $\omega$ meson to model the repulsion between nucleons. In order to take the isovector dependence of the interaction into account, an iso-vector, Lorentz-scalar $\delta$ meson and an iso-vector, Lorentz-vector $\rho$ meson are also considered. The notation of these fields follows their properties, i.e., $\sigma$, $\omega^{\mu}$, $\vec{\delta}$, and $\vec{\rho}^{\mu}$ with Lorentz indices $\mu$ and the vector symbol for the three isovector components. The pion as another important meson is not included as it does not appear in the final equations after applying the basic approximations of the approach. Besides the above mentioned mesons, the photon field, denoted by the vector $A^{\mu}$, completes the number of degrees of freedom to include also the electromagnetic interaction in the model.

The derivation of an EDF in a relativistic approach follows several steps that are detailed below. First, a Lagrangian density is formulated containing the contributions for the free propagation of the fields and their interaction. From it, the coupled field equations are derived that have to be solved self-consistently. This can be achieved in certain approximations and expressions for the energy of the system and other quantities of interest can be obtained. They are functions of the particle fields and their densities, so that the EDF is a generalized form of a density functional. It depends on a rather small number of parameters that need to be determined by appropriate procedures.
In the formulas given below, we use a system of units
where $\hbar = c = 1$. The value $\hbar = 197.3269804$~MeV~fm/c
is used for the conversion of units.

\subsection{Lagrangian density, field equations, and energy density}
\label{sec:LD_MFE}

The development of the relativistic EDF starts from a
covariant Lagrangian density
\begin{eqnarray}
\label{eq:L}
\nonumber \mathcal{L} & = & 
\sum_{\eta=\pi,\nu} \overline{\Psi}_{\eta}
\left( \gamma_{\mu}i\mathcal{D}^{\mu} - \sigma_{\mu\lambda}\mathcal{T}^{\mu\lambda}
- \mathcal{M}_{\eta} \right) \Psi_{\eta}
\\ \nonumber & &
+ \frac{1}{2} \Big( 
 \partial^{\mu} \sigma \partial_{\mu} \sigma - M_{\sigma}^{2} \: \sigma^{2}
 + \partial^{\mu} \vec{\delta} \cdot \partial_{\mu} \vec{\delta} 
 - M_{\delta}^{2} \: \vec{\delta} \cdot \vec{\delta}
 \\  & & 
 - \frac{1}{2} G^{\mu\lambda} G_{\mu\lambda}
 + M_{\omega}^{2} \: \omega^{\mu}\omega_{\mu}
 - \frac{1}{2} \vec{H}^{\mu\lambda} \cdot \vec{H}_{\mu\lambda}
 + M_{\rho}^{2} \: \vec{\rho}^{\mu} \cdot \vec{\rho}_{\mu}
  - \frac{1}{2} F^{\mu\lambda} F_{\mu\lambda}
 \Big) \: ,
\end{eqnarray}
with several contributions. Here, the field tensors of the mesons and of the
massless electromagnetic field are given by the usual expressions;
\begin{equation}
  G^{\mu\lambda}  =  \partial^{\mu} \omega^{\lambda}
  - \partial^{\lambda} \omega^{\mu} \: ,
  \qquad
  \vec{H}^{\mu\lambda}  =  \partial^{\mu} \vec{\rho}^{\lambda}
  - \partial^{\lambda} \vec{\rho}^{\mu} \: ,
  \qquad
  F^{\mu\lambda}  =  \partial^{\mu} A^{\lambda} - \partial^{\lambda} A^{\mu} \: .
\end{equation}
The minimal coupling to the nucleons
is realized by the corresponding terms in the covariant derivative
\begin{equation}
  \label{eq:D_cpl}
  i\mathcal{D}^{\mu} = i \partial^{\mu} - \tilde{\Gamma}_{\omega} \omega^{\mu}
  - \tilde{\Gamma}_{\rho} \vec{\rho}^{\mu} \cdot \vec{\tau}
  - \Gamma_{\gamma}  A^{\mu} \frac{1+\tau_{3}}{2} \: , 
\end{equation}
for the Lorentz-vector mesons and the photon, and in the mass operator
\begin{equation}
  \label{eq:M_cpl}
 \mathcal{M}_{\eta} =  M_{\eta} - \tilde{\Gamma}_{\sigma} \sigma 
 - \tilde{\Gamma}_{\delta} \vec{\delta} \cdot \vec{\tau} \: ,
\end{equation}
for the Lorentz-scalar mesons, respectively.
The term in Eq.\ (\ref{eq:L}) with the quantity
\begin{equation}
  \label{eq:T_cpl}
  \mathcal{T}^{\mu\lambda} = \frac{\Gamma_{T\omega}}{2M_{\pi}} G^{\mu\lambda}
  + \frac{\Gamma_{T\rho}}{2M_{\pi}} \vec{H}^{\mu\lambda} \cdot \vec{\tau}
  \: ,
\end{equation}
accounts for the additional tensor coupling of the nucleons to the mesons.
Tensor couplings to the electromagnetic field are not included.
The masses of the nucleons and the mesons are denoted
by $M_{\eta}$ ($\eta=\pi,\nu$) and $M_{j}$ ($j=\sigma,\delta,\omega,\rho$),
respectively.
The isospin dependence is expressed with help of the
isopin matrices $\tau_{k}$ ($k=1,2,3$)
in analogy to the Pauli spin matrices $\sigma_{k}$.
The coupling factors $\tilde{\Gamma}_{j}$
in Eqs.\ (\ref{eq:D_cpl}), and (\ref{eq:M_cpl})
are assumed to be functionals of the nucleon fields $\Psi_{\eta}$, and
$\overline{\Psi}_{\eta}= \Psi_{\eta}^{\dagger}\gamma^{0}$. 
In contrast,
the tensor couplings $\Gamma_{T\omega}$, and $\Gamma_{T\rho}$,
in Eq.\ (\ref{eq:T_cpl}) as well as
the electromagnetic coupling $\Gamma_{\gamma}$
in Eq.\ (\ref{eq:D_cpl}) are simple constants.
The devision by the proton mass in Eq.\ (\ref{eq:T_cpl})
is introduced to have dimensionless
couplings $\Gamma_{T\omega}$ and $\Gamma_{T\rho}$.

The field equations of the nucleons, mesons, and the photon
can be derived in the usual way from the Euler-Lagrange equations.
However, they cannot be solved in the field-theoretical sense and
approximations are needed to arrive at tractable equations.
The fields of the mesons and the photons are not quantized but considered
as classical fields and will be denoted by the same symbols as in the
original Lagrangian density (\ref{eq:L}) in the following. 
For the nucleons, the Hartree approximation is applied, i.e.,
the many-nucleon wave function $\Psi$ of the system of interest
is written as a product of single-nucleon states $\Psi_{\eta k_{\eta}m_{\eta}}$,
\begin{equation}
  \Psi = \prod_{\eta=\pi,\nu} \prod_{k_{\eta},m_{\eta}}
  w_{\eta k_{\eta}m_{\eta}} \Psi_{\eta k_{\eta}m_{\eta}} \: ,
\end{equation}
with occupation factors $w_{\eta k_{\eta}m_{\eta}}$.
Only values of $w_{\eta k_{\eta}m_{\eta}}=0$ and $1$ are allowed.
This condition reflects the fermionic nature of the nucleons.
The meaning of the quantum numbers $k_{\eta}$, and $m_{\eta}$, is explained
in Subsection \ref{sec:nuc_sph}.
Only single-nucleon states of positive-energy are considered here,
corresponding to the no-sea approximation
and only a finite number of states contributes in the
many-nucleon wave function.

For the dependence of the couplings $\tilde{\Gamma}_{j}$
in Eqs.\ (\ref{eq:D_cpl}), and (\ref{eq:M_cpl}), a so-called
vector density dependence is assumed, i.e., with the approximation
mentioned above, they become functions $\Gamma_{j}$ of the vector density,
\begin{equation}
  \label{eq:rho_v}
  \varrho_{v} = \sqrt{j_{\mu}j^{\mu}} \: ,
\end{equation}
with the total current,
\begin{equation}
  \label{eq:curr}
  j^{\mu} = \sum_{\eta=\pi,\nu} \sum_{k_{\eta},m_{\eta}}
  w_{\eta k_{\eta}m_{\eta}} \:
  \overline{\Psi}_{\eta k_{\eta}m_{\eta}} \gamma^{\mu} \Psi_{\eta k_{\eta}m_{\eta}}
  \: ,
\end{equation}
of the nucleons. The quantity $\varrho_{v}$ is in fact a Lorentz scalar
but has to be distinguished from the actual scalar density
\begin{equation}
  \label{eq:rho_s}
  \varrho_{s} = \sum_{\eta=\pi,\nu} \sum_{k_{\eta},m_{\eta} }
  w_{\eta k_{\eta}m_{\eta}} \:
  \overline{\Psi}_{\eta k_{\eta}m_{\eta}}  \Psi_{\eta k_{\eta}m_{\eta}}  \: ,
\end{equation}
that is used for couplings with a scalar density dependence,
c.f.\ Ref.\ \cite{Typel:2020ozc}.

In the actual application to nuclei and nuclear matter, a specific frame
of reference is chosen and the covariance of the equations will be lost.
In particular, a frame is chosen where the system is at rest,
the couplings depend only on the zero-component of the current
$\varrho_{v} = j^{0}$, and $\vec{j}=0$.
Since only static solutions without a change of isospin are considered,
only the zero-component of the Lorentz-vector fields and
a single component of the isovector quantities have to be taken
into account, since all other components vanish. 
Assuming spherical symmetry, the field equation of the nucleons assumes
the form of a modified Dirac equation,
\begin{equation}
  \label{eq:Dirac}
  \hat{H}_{\eta} \Psi_{\eta k_{\eta}m_{\eta}}(\vec{r})
  = \left[ \vec{\alpha} \cdot \hat{\vec{p}}
    + \beta \left( M_{\eta} - S_{\eta} \right)
    + V_{\eta} + i \vec{\gamma} \cdot \frac{\vec{r}}{r} T_{\eta} \right]
  \Psi_{\eta k_{\eta}m_{\eta}}(\vec{r})
  = E_{\eta k_{\eta}} \Psi_{\eta k_{\eta}m_{\eta}}(\vec{r}) \: ,
\end{equation}
with $\vec{\alpha}=\gamma^{0}\vec{\gamma}$, and $\beta=\gamma^{0}$,
for single-nucleon states of energy $E_{\eta k_{\eta}}$.
The interaction with the
meson and photon fields is expressed through the scalar potentials
\begin{equation}
\label{eq:S}
 S_{\eta}(\vec{r})  =  \Gamma_{\sigma} \sigma
 + g_{\eta}\Gamma_{\delta} \delta 
 \: ,
\end{equation}
vector potentials,
\begin{equation}
\label{eq:V}
 V_{\eta}(\vec{r})  =  \Gamma_{\omega} \omega_{0}
 + g_{\eta}\Gamma_{\rho} \rho_{0} 
 + \frac{1+g_{\eta}}{2} \Gamma_{\gamma} A_{0} + V^{(R)}
 \: ,
\end{equation}
and tensor potentials,
\begin{equation}
  \label{eq:T}
  T_{\eta}(\vec{r}) = - \frac{\Gamma_{T\omega}}{M_{\pi}} \frac{\vec{r}}{r}
  \cdot \vec{\nabla} \omega_{0}
  - g_{\eta}\frac{\Gamma_{T\rho}}{M_{\pi}} \frac{\vec{r}}{r}
  \cdot \vec{\nabla} \rho_{0} \: ,
\end{equation}
with factors $g_{\pi}=1$, and $g_{\nu}=-1$.
The vector density dependence of the couplings generates the
rearrangement contribution
\begin{equation}
  \label{eq:VR}
  V^{(R)}(\vec{r}) =
  \frac{d \Gamma_{\omega}}{d n^{(v)}} n_{\omega} \omega_{0}
 + \frac{d \Gamma_{\rho}}{d n^{(v)}}  n_{\rho} \rho_{0}
 -  \frac{d \Gamma_{\sigma}}{d n^{(v)}} n_{\sigma} \sigma
 -  \frac{d \Gamma_{\delta}}{d n^{(v)}} n_{\delta} \delta \: ,
\end{equation}
in the vector potential (\ref{eq:V}). It contains derivatives of the
couplings with respect
to the vector density $n^{(v)}=\varrho_{v}$, as well as meson fields and
source densities $n_{j}$ ($j=\sigma,\delta,\omega,\rho$)
that also appear in the meson field equations;
\begin{eqnarray}
  \label{eq:sigma}
  -\Delta \sigma + M_{\sigma}^{2} \sigma
  & = &  \Gamma_{\sigma} n_{\sigma} \: ,
  \\
  \label{eq:delta}
  -\Delta \delta + M_{\delta}^{2} \delta
  & = &  \Gamma_{\delta} n_{\delta}\: ,
  \\
  \label{eq:omega}
  -\Delta \omega_{0} + M_{\omega}^{2} \omega_{0}
  & = &  \Gamma_{\omega} n_{\omega}
  + \frac{\Gamma_{T\omega}}{m_{p}} \vec{\nabla} \cdot \vec{\jmath}^{(t)}_{\omega}
  \: ,
  \\
  \label{eq:rho}
  -\Delta \rho_{0} + M_{\rho}^{2} \rho_{0}
  & = &  \Gamma_{\rho} n_{\rho}
  + \frac{\Gamma_{T\rho}}{m_{p}} \vec{\nabla} \cdot \vec{\jmath}^{(t)}_{\rho}
  \: ,
\end{eqnarray}
for the non-vanishing components of the fields.
They are given by
\begin{eqnarray}
  & & n_{\sigma} = n^{(s)} = n_{\pi}^{(s)}+n_{\nu}^{(s)} \: ,\qquad
  n_{\delta} = n_{\pi}^{(s)}-n_{\nu}^{(s)} \: ,
  \\ & & n_{\omega} = n^{(v)} = n_{\pi}^{(v)}+n_{\nu}^{(v)} \: , \qquad
  n_{\rho} = n_{\pi}^{(v)}-n_{\nu}^{(v)} \: ,
\end{eqnarray}
with the vector and scalar densities;
\begin{equation}
  \label{eq:ns}
  n_{\eta}^{(v)}(\vec{r}) =
  \sum_{k_{\eta},m_{\eta}} w_{\eta k_{\eta}m_{\eta}} n_{\eta k_{\eta}m_{\eta}}^{(v)}
  = \sum_{k_{\eta},m_{\eta}} w_{\eta k_{\eta}m_{\eta}}
  \overline{\Psi}_{\eta k_{\eta}m_{\eta}} \gamma^{0} \Psi_{\eta k_{\eta}m_{\eta}} \: ,
\end{equation}
and
\begin{equation}
  \label{eq:nv}
  n_{\eta}^{(s)}(\vec{r}) =
  \sum_{k_{\eta},m_{\eta}} w_{\eta k_{\eta}m_{\eta}} n_{\eta k_{\eta}m_{\eta}}^{(s)}
  = \sum_{k_{\eta}m_{\eta}} w_{\eta k_{\eta}m_{\eta}}
  \overline{\Psi}_{\eta k_{\eta}m_{\eta}} \Psi_{\eta k_{\eta}m_{\eta}} \: ,
\end{equation}
of nucleons, respectively.
In Eqs.\ (\ref{eq:omega}), and (\ref{eq:rho}), also
the divergences of the tensor currents
\begin{equation}
   \label{eq:tcurrent}
   \vec{\jmath}^{(t)}_{\omega} = \vec{\jmath}_{\pi}^{(t)} + \vec{\jmath}_{\nu}^{(t)} \: ,
   \qquad
   \vec{\jmath}^{(t)}_{\rho} = \vec{\jmath}_{\pi}^{(t)} - \vec{\jmath}_{\nu}^{(t)} \: ,
\end{equation}
with,
\begin{equation}
  \label{eq:tcurrent_i}
  \vec{\jmath}_{\eta}^{(t)}(\vec{r}) =
  \sum_{k_{\eta},m_{\eta}} w_{\eta k_{\eta}m_{\eta}} \:
  \vec{\jmath}_{\eta k_{\eta}m_{\eta}}^{(t)}
  = \sum_{k_{\eta},m_{\eta}} w_{\eta k_{\eta}m_{\eta}} \:
  \overline{\Psi}_{\eta k_{\eta}m_{\eta}} i \vec{\alpha} \Psi_{\eta k_{\eta}m_{\eta}} \: ,
\end{equation}
contribute to the source terms of the Lorentz-vector meson fields
in Eqs.\ (\ref{eq:omega}), and (\ref{eq:rho}).
Note that these tensor currents are not the vector part of the usual
  four-currents,
  \begin{equation}
    j_{\eta}^{\mu}(\vec{r}) =
    \sum_{k_{\eta},m_{\eta}} w_{\eta k_{\eta}m_{\eta}} \:
  \overline{\Psi}_{\eta k_{\eta}m_{\eta}} \gamma^{\mu} \Psi_{\eta k_{\eta}m_{\eta}} \: .
\end{equation}
Finally, the field equation for the electromagnetic field is given by
the usual Poisson's equation
\begin{equation}
  \label{eq:gamma}
    -\Delta A_{0} =  \Gamma_{\gamma} n_{\gamma} \: ,
\end{equation}
with the source density $n_{\gamma}=n_{\pi}^{(v)}$.

In the application to nuclear matter, there is no spatial dependence
of the source densities and meson fields. They become constants and there
are no contributions from the tensor currents in Eqs.\ (\ref{eq:omega}),
and (\ref{eq:rho}). The field equations are solved trivially
and the tensor potential (\ref{eq:T}) in the Dirac equation
(\ref{eq:Dirac}) vanishes too.
Nucleon states are then described by plane waves.

The energy density of the system
can be obtained as a component of the energy-momentum tensor
\begin{equation}
  \label{eq:Tmunu}
  T^{\mu\lambda} = \sum_{\varphi}
  \frac{\partial \mathcal{L}}{\partial (\partial_{\mu} \varphi)}
  \partial^{\lambda} \varphi - g^{\mu\lambda} \mathcal{L} \: ,
\end{equation}
where the sum runs over all fields
$\varphi=\Psi_{\eta k_{\eta}m_{\eta}},\overline{\Psi}_{\eta k_{\eta}m_{\eta}},
\sigma,\delta,\omega_{0},\rho_{0},A_{0}$,
and the metric tensor is $g^{\mu\lambda}=\mbox{diag}(-1,1,1,1)$.
It can be expressed as a sum,
\begin{equation}
  \varepsilon(\vec{r}) = T^{00} =
  \varepsilon_{\mathrm{nucleon}}(\vec{r})
  + \varepsilon_{\mathrm{field}}(\vec{r}) \: ,
\end{equation}
with the contribution
\begin{equation}
  \varepsilon_{\mathrm{nucleon}}(\vec{r}) =
  \sum_{\eta=\pi,\nu} \sum_{k_{\eta},m_{\eta}}
  w_{\eta k_{\eta}m_{\eta}} E_{\eta k_{\eta}} n^{(v)}_{\eta k_{\eta}m_{\eta}} \: ,
\end{equation}
of the nucleons and that of the fields
\begin{eqnarray}
  \label{eq:eps_field}
 \nonumber  \varepsilon_{\mathrm{field}}(\vec{r}) & = &
 \frac{1}{2} \left( 
 \vec{\nabla} \sigma \cdot \vec{\nabla} \sigma + M_{\sigma}^{2} \sigma^{2}
 + \vec{\nabla} \delta \cdot \vec{\nabla} \delta + M_{\delta}^{2} \delta^{2}
  \right. \\  & & \left.
  - \vec{\nabla} \omega_{0} \cdot \vec{\nabla} \omega_{0}
  - M_{\omega}^{2} \omega_{0}^{2}
  - \vec{\nabla} \rho_{0} \cdot \vec{\nabla} \rho_{0} - M_{\rho}^{2} \rho_{0}^{2}
  - \vec{\nabla} A_{0} \cdot \vec{\nabla} A_{0}
 \right)
 - V^{(R)} n^{(v)} 
\end{eqnarray}
with an explicit rearrangement term.

\subsection{Spherical nuclei}
\label{sec:nuc_sph}

For describing nuclei, the coupled differential equations (\ref{eq:Dirac})
for the single-nucleon wave functions, and
(\ref{eq:sigma}) - (\ref{eq:gamma}), have to be solved self-consistently.
In the case of spherical symmetry, it is convenient to represent the
single-nucleon wave functions as
\begin{equation}
  \label{eq:Psi}
  \Psi_{\eta k_{\eta}m_{\eta}}(\vec{r}) = \frac{1}{r} \left( \begin{array}{l}
 F_{\eta k_{\eta}}(r) \mathcal{Y}_{\kappa_{\eta}m_{\eta}}(\hat{r}) \\
 i G_{\eta k_{\eta}}(r) \mathcal{Y}_{-\kappa_{\eta}m_{\eta}}(\hat{r})
   \end{array}\right) \: ,
\end{equation}
with radial functions $F_{\eta k_{\eta}}$, and $G_{\eta k_{\eta}}$,
and tensor spherical harmonics,
\begin{equation}
  \mathcal{Y}_{\kappa_{\eta} m}(\hat{r}) = \sum_{m_{l} m_{s}}
  (l_{\eta} \: m_{l} \: s \: m_{s} | j_{\eta} \: m_{\eta})
  Y_{l_{\eta} m_{l}}(\hat{r}) \chi_{sm_{s}}  \: ,
\end{equation}
in the upper and lower components of the Dirac spinor.
Each state is identified by the index $\eta=\pi,\nu$, and two quantum numbers
$n_{\eta}$, and $\kappa_{\eta}$, that are collectively denoted as $k_{\eta}$, and
a total angular momentum projection $m_{\eta}$.
The first, $n_{\eta}=0,1,2,\dots$,
is the principal quantum number related to the number of
nodes of the function $F_{\eta k_{\eta}}$;
the second, $\kappa_{\eta}=\pm 1, \pm 2,\dots$, defines
the total angular momentum
$j_{\eta}=|\kappa_{\eta}|-1/2$
and the orbital angular momentum
$l_{\eta}=j_{\eta}-\kappa_{\eta}/(2|\kappa_{\eta}|)$
of a state.
Then the Dirac equation (\ref{eq:Dirac}) is equivalent to the
set of two coupled differential equations;
\begin{equation}
  \label{eq:deq_F}
  \left( E_{\eta k_{\eta}} - V_{\eta} - M_{\eta} + S_{\eta} \right)
  F_{\eta k_{\eta}}  
  +\left( \frac{d}{dr} + \frac{\kappa_{\eta}}{r}
  - T_{\eta} \right)
  G_{\eta k_{\eta}}  =  0 \: ,
\end{equation}
and,
\begin{equation}
  \label{eq:deq_G}
  \left( \frac{d}{dr} - \frac{\kappa_{\eta}}{r}
    + T_{\eta} \right)
  F_{\eta k_{\eta}}
  - \left( E_{\eta k_{\eta}} - V_{\eta} + M_{\eta} - S_{\eta} \right)
  G_{\eta k_{\eta}}  =  0 \: , 
\end{equation}
for the radial functions that have to be solved with the boundary conditions
$F_{\eta k_{\eta}}(0)=G_{\eta k_{\eta}}(0)=0$, and
$\lim_{r \to 0}F_{\eta k_{\eta}}(r)= \lim_{r \to 0}G_{\eta k_{\eta}}(r) = 0$.

The representation (\ref{eq:Psi}) of the single-nucleon wavefunctions leads
to the expressions
\begin{equation}
  \label{eq:ni_s}
  n_{\eta k_{\eta}m_{\eta}}^{(s)}(r)  =  \frac{1}{4\pi r^{2}}
  \left[ \left| F_{\eta k_{\eta}}(r)\right|^{2}
    - \left| G_{\eta k_{\eta}}(r)\right|^{2} \right] 
\end{equation}
and
\begin{equation}
  \label{eq:ni_v}
 n_{\eta k_{\eta}m_{\eta}}^{(v)}(r)  =  \frac{1}{4\pi r^{2}}
 \left[ \left| F_{\eta k_{\eta}}(r)\right|^{2}
   + \left| G_{\eta k_{\eta}}(r)\right|^{2} \right]
\end{equation}
for the scalar and vector single-nucleon densities
in Eqs.\ (\ref{eq:nv}), and (\ref{eq:ns}), respectively,
if an average over the quantum number $m_{\eta}$ is applied.
This procedure is used to obtain spherical source densities
in the field equations for the mesons and the 
photon in case that sub-shells of given $j$ are not completely filled.
The vector densities are normalized as,
\begin{equation}
 \int d^{3}r \: n_{\eta k_{\eta}m_{\eta}}^{(v)}(r)  = 1 \: ,
\end{equation}
but, there is no such condition for the scalar densities.
The single-nucleon tensor currents have only
a component in radial direction in a spherical system. Then one can define
\begin{equation}
 n_{\eta k_{\eta}m_{\eta}}^{(t)}  =  \frac{1}{4\pi r^{2}}
 \left[ F_{\eta k_{\eta}}^{\ast}(r) G_{\eta k_{\eta}}(r)
   + F_{\eta k_{\eta}}(r) G_{\eta k_{\eta}}^{\ast}(r) \right]  \: ,
\end{equation}
so that,
\begin{equation}
  \vec{\jmath}_{\eta k_{\eta}m_{\eta}}^{(t)}
  = n_{\eta k_{\eta}m_{\eta}}^{(t)} \frac{\vec{r}}{r} \: ,
  \qquad \mbox{and} \qquad
  \vec{\nabla} \cdot \vec{\jmath}_{\eta k_{\eta}m_{\eta}}^{(t)}
  = \frac{1}{r^{2}} \frac{d}{dr}
  \left( r^{2} n_{\eta k_{\eta}m_{\eta}}^{(t)}\right) \: ,
\end{equation}
in Eq.\ (\ref{eq:tcurrent_i}) after applying the averaging procedure.
The occupation factors in the calculation of the densities are determined
such that,
\begin{equation}
  \sum_{k_{\pi},m_{\pi}} w_{\pi k_{\pi}m_{\pi}} = Z \: , \qquad \mbox{and} \qquad
  \sum_{k_{\nu}m_{\nu}} w_{\nu k_{\nu}m_{\nu}} = N \: , 
\end{equation}
with the proton and neutron numbers $Z$ and $N$ of the nucleus,
respectively, with mass number $A=Z+N$.

The energy of the nucleus is finally obtained as
\begin{equation}
  \label{eq:E_nuc}
  E_{\mathrm{nucleus}} = E_{\mathrm{sp}} + E_{\mathrm{field}} - E_{\mathrm{corr}} \: ,
\end{equation}
with the contributions of the single-particle states
\begin{equation}
  E_{\mathrm{sp}} = \sum_{\eta=\pi,\nu}
  \sum_{k_{\eta},m_{\eta}} w_{\eta k_{\eta}m_{\eta}} E_{\eta k_{\eta}} \: , 
\end{equation}
and of the fields, 
\begin{eqnarray}
  \nonumber E_{\mathrm{field}} & = &
  \frac{1}{2} \int d^{3}r \:
  \Big\{ \Gamma_{\sigma} n_{\sigma} \sigma
  + \Gamma_{\delta} n_{\delta} \delta
  - \Gamma_{\gamma} n_{\gamma} A_{0}
  \\   & &
    - \left[ \Gamma_{\omega} n_{\omega}
    + \frac{\Gamma_{T\omega}}{M_{p}} \vec{\nabla} \cdot \vec{\jmath}_{T\omega}
    \right] \omega_{0}
  - \left[ \Gamma_{\rho} n_{\rho}
    + \frac{\Gamma_{T\rho}}{M_{p}} \vec{\nabla} \cdot \vec{\jmath}_{T\rho}
    \right] \rho_{0}
   - V^{(R)} n^{(v)} 
  \Big\} \: , 
\end{eqnarray}
where the field equations were exploited after a partial integration
using Eq.\ (\ref{eq:eps_field}). Since the translational symmetry is broken,
a correction for the center-of-mass motion has to be applied. It is calculated
as the expectation value
\begin{equation}
  \label{eq:E_corr}
  E_{\mathrm{corr}} = \langle \Psi | \frac{\vec{P}^{2}}{2M(Z,N)} | \Psi \rangle  \: , 
\end{equation}
of the total momentum
\begin{equation}
  \vec{P}= \sum_{j=1,Z} \vec{p}_{\pi j} + \sum_{j=1,N} \vec{p}_{\nu j} \: , 
\end{equation}
in non-relativistic form
with the total mass $M(Z,N)=ZM_{\pi}+NM_{\nu}$ of the nucleus,
neglecting the binding energy.

For the numerical solution of the coupled differential equations
(\ref{eq:deq_F}), and (\ref{eq:deq_G}), the Lagrange-mesh technique
is utilized, see Ref.\ \cite{Typel:2018tvg} and references therein.  
This approach combines the virtues of an expansion in terms of
basis functions and a discretization on a grid. At the same time
it is numerically efficient and precise.
The radial wave functions are represented with their values at grid points
$r_{i}=h x_{i}^{(\alpha)}$, where $h$ is a scale parameter, and
$x_{i}^{(\alpha)}$ are the zeroes
$i=1,\dots,N_{\mathrm{zero}}$ of the Laguerre
polynomials $L_{N}^{(\alpha)}(x)$ with $N_{\mathrm{zero}}=25$,
and $\alpha=2|\kappa|$.
The scale parameter $h$ is chosen such that the largest zero
of $L_{N}^{(0)}$ is at a radius $r=0.95R$, with $R=r_{0} A^{1/3}+10$~fm, and
$r_{0}=1.2$~fm depending
on the mass number $A$ of the nucleus. The field equations of the mesons,
(\ref{eq:sigma}) - (\ref{eq:rho}),
and the photon, (\ref{eq:gamma}),
are solved by expanding the fields and sources in terms of
Riccati-Bessel functions. Special care has to be taken for the electromagnetic
potential $A_{0}$ to obtain the correct asymtotic behavior
for large radii. For more details on the numerical techniques,
see Ref.\ \cite{Typel:2018tvg}.

\subsection{Nuclear matter}

The solution of the field equations
(\ref{eq:Dirac}), (\ref{eq:sigma}) - (\ref{eq:rho}), and (\ref{eq:gamma})
simplifies considerably in homogeneous and isotropic nuclear matter.
The source densities and meson fields become constants. The electromagnetic
field has to be discarded and all contributions of the tensor terms vanish. 
Nucleons are described by plane waves of momentum $k_{\eta}$
following the dispersion relation
\begin{equation}
  E_{\eta k_{\eta}} = \sqrt{k_{\eta}^{2}+\left( M_{\eta}^{\ast}\right)^{2}} + V_{\eta}
  \: ,
\end{equation}
where the Dirac effective mass
\begin{equation}
  M_{\eta}^{\ast} = M_{\eta} - S_{\eta} \: ,
\end{equation}
and constant scalar and vector potentials (\ref{eq:S}), and (\ref{eq:V}), appear.
At zero temperature, all relevant quantities can be calculated analytically.
The scalar and vector densities assume the form;
\begin{equation}
  \label{eq:nsi}
  n_{\eta}^{(s)} = \gamma_{\eta} \int_{0}^{k_{\eta}^{\ast}} \frac{d^{3}k_{\eta}}{(2\pi)^{3}}
  \: \frac{M_{\eta}^{\ast}}{\sqrt{\left(k_{\eta}\right)^{2}
      +\left(M_{\eta}^{\ast}\right)^{2}}}
   =   \frac{\gamma_{\eta}M_{\eta}^{\ast}}{4\pi^{2}} \left[ k_{\eta}^{\ast}\mu_{\eta}^{\ast}
    - \left( M_{\eta}^{\ast} \right)^{2}
    \ln \frac{k_{\eta}^{\ast}+\mu_{\eta}^{\ast}}{M_{\eta}^{\ast}} \right] \: ,
\end{equation}
and,
\begin{equation}
  \label{eq:nvi}
  n_{\eta}^{(v)} = \gamma_{\eta} \int_{0}^{k_{\eta}^{\ast}} \frac{d^{3}k_{\eta}}{(2\pi)^{3}} =
  \frac{g_{\eta}}{6\pi^{2}} \left(k_{\eta}^{\ast}\right)^{3} \: ,
\end{equation}
in the continuum approximation, with the spin degeneracy factor $\gamma_{\eta}=2$.
The Fermi momenta $k_{\eta}^{\ast}$ are related to the
effective chemical potentials 
\begin{equation}
  \label{eq:mu_eff}
  \mu_{\eta}^{\ast} = \mu_{\eta} - V_{\eta}
  = \sqrt{\left(k_{\eta}^{\ast}\right)^{2}+\left( M_{\eta}^{\ast}\right)^{2}}  \: ,
\end{equation}
of the nucleons.

The energy density of cold nuclear matter can be written as,
\begin{equation}
  \label{eq:eps}
  \varepsilon = \sum_{\eta=\pi,\nu} 
  \frac{1}{4} \left[ 3\mu_{\eta}^{\ast}n_{\eta}^{(v)}
    + M_{\eta}^{\ast}n_{\eta}^{(s)} \right]
  + \frac{1}{2} \sum_{j=\sigma,\delta,\omega,\rho} C_{j} n_{j}^{2} \: ,
\end{equation}
with the coupling-to-mass ratios 
\begin{equation}
  C_{j} = \frac{\Gamma_{j}^{2}}{M_{j}^{2}} \: .
\end{equation}
The pressure
\begin{equation}
  \label{eq:pres}
  P = \sum_{\eta=\pi,\nu} 
  \frac{1}{4} \left[ \mu_{\eta}^{\ast}n_{\eta}^{(v)}
    - M_{\eta}^{\ast}n_{\eta}^{(s)} \right] 
  - \frac{1}{2} \sum_{j=\sigma,\delta}
  \left( C_{j} + n^{(v)} C_{j}^{\prime} \right) n_{j}^{2}
  + \frac{1}{2} \sum_{j=\omega,\rho}
  \left( C_{j} + n^{(v)} C_{j}^{\prime}\right) n_{j}^{2} \: , 
\end{equation}
also depends on the derivatives
\begin{equation}
  C_{j}^{\prime} = \frac{dC_{j}}{dn^{(v)}} \: ,
\end{equation}
and can be derived using the thermodynamic definition
$P = d(\varepsilon/n^{(v)})/dn^{(v)}$, or from the energy-momentum tensor
(\ref{eq:Tmunu}), as $P = (1/3)\sum_{\mu=1}^{3}T^{\mu\mu}$.
It is easily shown that energy density and pressure
fulfill the Hugenberg-van-Hove relation
\cite{Hugenholtz:1958zz}
\begin{equation}
  \varepsilon + P = \sum_{\eta=\pi,\nu} \mu_{\eta} n_{\eta}^{(v)} \: ,
\end{equation}
since the rearrangement terms in the pressure combine with those
in the chemical potentials $\mu_{\eta}$ appearing in Eq.\ (\ref{eq:mu_eff}).

The energy density (\ref{eq:eps}), and pressure (\ref{eq:pres}), are functions
of two independent densities, the proton and neutron vector densities
$n_{\pi}^{(v)}$, and $n_{\nu}^{(v)}$.
Alternatively, they can be regarded
as functions of the baryon density
\begin{equation}
  n_{b} = \varrho_{v} = n^{(v)} = n_{\nu}^{(v)} + n_{\pi}^{(v)} \: ,
\end{equation}
and the neutron-proton asymmetry
\begin{equation}
  \alpha = \frac{n_{\nu}^{(v)} - n_{\pi}^{(v)}}{n_{b}} \: ,
\end{equation}
that lies inside the interval $[-1,+1]$. Then it is convenient to define
the energy per nucleon
\begin{equation}
  \label{eq:EA}
  \frac{E}{A}(n_{b},\alpha) = \frac{\varepsilon}{n_{b}} - M_{\mathrm{nuc}} \: ,
\end{equation}
substracting the average nucleon mass $M_{\mathrm{nuc}}=(M_{\nu}+M_{\pi})/2$.
A second derivative,
\begin{equation}
  \label{eq:Esym}
  E_{\mathrm{sym}}(n_{b}) =
  \frac{1}{2} \left. \frac{d^{2}}{d\alpha^{2}}
  \frac{E}{A} \right|_{\alpha=0} \: ,
\end{equation}
gives the symmetry energy.
It characterizes how the isospin dependence of the energy per nucleon
changes with the baryon density.

The energy per nucleon of symmetric matter
near saturation can be expanded as
\begin{equation}
  E_{0}(n_{b}) = \frac{E}{A} (n_{b},0) = -B_{\mathrm{sat}}
  + \frac{1}{2} K x^{2} + \frac{1}{6} Q x^{3} + \dots \: ,
\end{equation}
with $x=(n_{b}-n_{\mathrm{sat}})/(3n_{\mathrm{sat}})$
without a term linear in $x$ because
the pressure at saturation is
$P(n_{\mathrm{sat}})=0$.
The incompressibility coefficient
\begin{equation}
  K = 9 n_{\rm sat}^{2}
  \left.  \frac{d^{2}}{d(n_{b})^{2}} E_{0} \right|_{n_{\mathrm{sat}}}
  = 9 \left.  \frac{dP}{dn_{b}} \right|_{n_{\mathrm{sat}}} \: , 
\end{equation}
and skewness parameter
\begin{equation}
  Q = 27 n_{\rm sat}^{3}
  \left.  \frac{d^{3}}{d(n_{b})^{3}} E_{0}
  \right|_{n_{\rm sat}} \: , 
\end{equation}
are found as second and third derivatives of $E_{0}$.
Similarly, the symmetry energy can be written as
\begin{equation}
  E_{\mathrm{sym}}(n_{b}) = J + L x  
  + \frac{1}{2} K_{\mathrm{sym}} x^{2} + \dots \: ,
\end{equation}
with the symmetry energy at saturation $J=E_{\mathrm{sym}}(n_{\mathrm{sat}})$,
the slope parameter
\begin{equation}
  L = 3 n_{\mathrm{sat}} \left. \frac{dE_{\mathrm{sym}}}{dn_{b}} \right|_{n_{\mathrm{sat}}} \: , 
\end{equation}
and the symmetry incompressibility
\begin{equation}
  K_{\rm sym} = 9 n_{\rm sat}^{2}
  \left.  \frac{d^{2} E_{\mathrm{sym}}}{d(n_{b})^{2}}  \right|_{n_{\mathrm{sat}}} \: .
\end{equation}
It is also useful to define the socalled volume part of the isospin
compressibility 
\begin{equation}
  K_{\tau,v} = K_{\rm sym}-6L-\frac{QL}{K} \: ,
\end{equation}
that describes the change of the incompressibility with the variation
in $\alpha$ \cite{Dutra:2014qga,Vidana:2009is}.
These quantities, together with the effective Dirac mass at saturation
$M_{\mathrm{sat}}^ {\ast} = M_{\mathrm{nuc}}^{\ast}(n_{\mathrm{sat}})=
M_{\mathrm{nuc}}-C_{\sigma}n_{\mathrm{sat}}^{(s)}$,
are convenient to characterize the equation of state of nuclear matter
around saturation.

\section{Determination of model parameters}
\label{sec:para}

The relativistic EDF depends on several parameters: the masses of the
particles and the coefficients that describe the strength of the
couplings and their density dependence.
The masses of the nucleons and the mesons, except that of the
$\sigma$ meson are usually chosen to be identical or close to the experimental
values. In the present application of the relativistic EDF
the $\delta$ meson is not included in the model.
The remaining fixed mass parameters of the nucleons and Lorentz-vector mesons
are given in Table \ref{tab:masses} for the models in this work.
The mass of the $\sigma$ meson is used as an independent parameter because
it is not well defined by experimental data and affects the surface
properties of nuclei.

\begin{table}[t]
  \caption{Masses of nucleons and Lorentz-vector mesons used
    in the model parameterisations of this work.}
  \label{tab:masses}
  \begin{tabular}{@{}lcccc@{}}
    \toprule
    particle & $p$ & $n$ & $\omega$ & $\rho$\\
    \midrule
    mass in MeV & $938.27208816$ & $939.5654205$ & $783$ & $763$ \\
    \bottomrule
  \end{tabular}
\end{table}

The actual values of the model parameters are determined by fitting selected
observables of finite nuclei that are sensitive to different contributions
in the EDF and are obtained more-or-less directly from experiment.
More indirectly obtained quantities,
e.g., characteristic nuclear matter parameters, that can depend
strongly on the extraction method from actual experimental data,
or other constraints, e.g.,  from ab-initio calculations
of nuclear matter or from astrophysical observations, are not used
in this procedure. A comparison to these data will be performed after
the model parameters have been determined.

\subsection{Functional form of density dependent couplings}

The meson-nucleon couplings $\Gamma_{j}$ 
are assumed to be functions
of the vector density $\varrho_{v}=n^{(v)}$. In this work, the functional
form of the dependence is chosen as introduced
originally in \cite{Typel:1999yq} with
\begin{equation}
  \label{eq:Gamma_dd}
  \Gamma_{j}(\varrho) = \Gamma_{j}^{(0)} f_{j}(x) \: ,
\end{equation}
where $\Gamma_{j}^{(0)}=\Gamma_{j}(\varrho_{\mathrm{ref}})$
is the coupling at a reference density
$\varrho_{\mathrm{ref}}$ and the functions $f_{j}$ depend on the
parameter $x=\varrho_{v}/\varrho_{\mathrm{ref}} \geq 0$.
For the isoscalar mesons $j=\sigma,\omega$ a rational form
\begin{equation}
  \label{eq:fso}
  f_{j}(x) = a_{j} \frac{1+b_{j}(x+d_{j})^{2}}{1+c_{j}(x+d_{j})^{2}} \: , 
\end{equation}
with four coefficients $a_{j}$, $b_{j}$, $c_{j}$, and $d_{j}$ is used.
To avoid a pole in $f_{j}$, the coefficients
$c_{j}$, and $d_{j}$ have to be non-negative.
The constraints $f_{j}(1)=1$, and $f_{j}^{\prime\prime}(0)=0$,
reduce the number of independent coefficients to two with
$a_{j}=[1+c_{j}(1+d_{j})^{2}]/[1+b_{j}(1+d_{j})^{2}]$,
and $c_{j}=1/(3d_{j}^{2})$. The coupling function for the isovector mesons
$j=\rho$ is chosen as an exponential,
\begin{equation}
  \label{eq:frho}
  f_{\rho}(x) = \exp[-a_{\rho}(x-1)] \: ,
\end{equation}
with only one coefficient $a_{\rho}$.
Obviouly, other functions can be and have been used
to describe the density dependence of the couplings, see, e.g., Ref.\
\cite{Dutra:2014qga}.
Also a dependence on the scalar density $n^{(s)}$ or other densities
can be explored, see, e.g., Ref.\ \cite{Typel:2020ozc}.

In the model parameterisations of this work,
there are three independent parameters for the isoscalar mesons
$\sigma$ and for $\omega$
and two for the isovector meson $\rho$,
thus eight parameters in total that need to be determined
for the nuclear-matter calculations besides the mass of the $\sigma$ meson.
In the actual fitting procedure, the saturation density of symmetric nuclear
matter $n_{\mathrm{sat}}$ is used as reference density $\varrho_{\mathrm{ref}}$,
but the couplings $\Gamma_{j}^{(0)}$
at $\varrho_{\mathrm{ref}}$ and the coefficients
$a_{j}$, $b_{j}$, $c_{j}$, and $d_{j}$ are not used directly as independent
quantities that are varied.
Instead, the following characteristic nuclear
matter parameters are selected as variable quantities:
\begin{enumerate}
\item the effective Dirac mass at saturation $M_{\mathrm{sat}}^{\ast}$,
\item the binding energy per nucleon at saturation
  $B_{\mathrm{sat}}$, 
\item the incompressibility coefficient at saturation $K$,
\item the symmetry energy at saturation $J$,
\item the slope parameter of the symmetry energy at saturation $L$.
\end{enumerate}
Together with the condition of vanishing pressure at $n_{\mathrm{sat}}$,
this gives already six parameters. 
The remaining two quantities
that are used in the actual fit are the ratios
\begin{equation}
  \label{eq:ratios}
  r_{10} = \frac{f_{\omega}^{\prime}(1)}{f_{\omega}(1)} \: ,
  \qquad \mbox{and} \qquad
  r_{21} = \frac{f_{\omega}^{\prime\prime}(1)}{f_{\omega}^{\prime}(1)} \: , 
\end{equation}
with the density dependent function $f_{\omega}(x)$ of the $\omega$ meson.
It is assumed that the coupling functions are decreasing
with the density, hence $f_{\omega}^{\prime}(1)$ has to be smaller than zero.
Then the functional form of $f_{\omega}$ leads to the constraint
$-3 < r_{21} < 0$ 
for the second ratio in Eq.\ (\ref{eq:ratios}).
The actual procedure of the parameter conversion is given in detail
in Appendix \ref{sec:appA}.

In the description of finite nuclei, two additional tensor
couplings $\Gamma_{T\omega}$ and $\Gamma_{T\rho}$ appear if the tensor
terms in the Lagrangian density (\ref{eq:L}) are considered. These constants
are used directly as independent parameters in the fit.
Hence there are in total 11 (9) independent parameters in models with (without)
tensor couplings taking into account the mass of the $\sigma$ meson
$M_{\sigma}$ as another undetermined quantity.

\subsection{Observables and nuclei in the fit}
\label{sec:obs}

The choice of the observables that are included in fitting the model
parameters will affect the final results. Only quantities should be
selected that are well determined by experiments. Different
strategies are followed by various authors.
In the present work, the following nuclear observables of a nucleus are used:
\begin{enumerate}
\item the binding energy $B$,
\item the charge radius $r_{c}$,
\item the diffraction radius $r_{d}$,
\item the surface thickness $\sigma$,
\item the root-mean-square radius $r_{n}$ of a single valence
  neutron level above a shell closure,
  \item the spin-orbit splitting $\Delta E_{so}$
  for two levels of a nucleon $\eta$
  in states with the same quantum numbers $n_{\eta}$ and $l_{\eta}$ but different
  $j_{\eta}=l_{\eta}\pm 1/2$,
\item the constraint isoscalar monopole giant resonance energy $E_{\mathrm{mono}}$.
\end{enumerate}
They are calculated after the coupled equations for a nucleus $(Z,N)$
are solved and can be compared with experimental data.

\subsubsection{Values of observables from theory}

The binding energy is derived as
\begin{equation}
  B(Z,N) = M(Z,N)-E_{\mathrm{nucleus}}(Z,N)  \: ,
\end{equation}
from the energy (\ref{eq:E_nuc}).
The charge radius, the diffraction radius, and the surface thickness
are obtained from the charge form factor of a spherical nucleus. It depends
on a momentum $\vec{q}$
and is calculated as,
\begin{equation}
  \label{eq:F_ZN}
  F_{Z,N}(q) = \left[F_{Z}(q)F_{\pi}(q)
    +F_{N}(q)F_{\nu}(q)\right] F_{\mathrm{corr}}(q) \: ,
\end{equation}
with form factors of the proton and neutron distributions in the nucleus,
single-nucleon form factors 
and a center-of-mass correction.
The density distributions (\ref{eq:ni_v}) of neutron and protons are
used to calculate the form factors
\begin{equation}
  F_{Z}(q) = \int d^{3}r \:
  \sum_{k_{\pi},m_{\pi}} w_{\pi k_{\pi}m_{\pi}} n_{\pi k_{\pi}m_{\pi}}(\vec{r}) \:
  \exp \left( i\vec{q} \cdot \vec{r} \right) \: ,
\end{equation}
and,
\begin{equation}
  F_{N}(q)= \int d^{3}r \:
  \sum_{k_{\nu},m_{\nu}} w_{\nu k_{\nu}m_{\nu}} n_{\nu k_{\nu}m_{\nu}}(\vec{r}) \:
  \exp \left( i\vec{q} \cdot \vec{r} \right) \: ,
\end{equation}
that are spherically symmetric in momentum space due to the spherical
symmetry of the density distributions.
The single-nucleon form factors are parameterized
\cite{Simon:1980hu,Reinhard:1986qq} as
\begin{equation}
  \label{eq:F_pn}
  F_{\eta}(q) = \sum_{k=1}^{4} \frac{A_{\eta k}}{1+B_{\eta k}q^{2}} \: ,
\end{equation}
with coefficients $A_{\eta k}$, and $B_{\eta k}$,
that are given in Table \ref{tab:F_pn}.
The center-of-mass correction
\begin{equation}
  F_{\mathrm{corr}}(q) =
  \exp \left[\frac{3q^{2}}{16E_{\mathrm{corr}}M(Z,N)}\right] \: ,
\end{equation}
depends on the center-of-mass correction (\ref{eq:E_corr}) of the energy.
The charge density of the nucleus is finally obtained from
\begin{equation}
  n_{\mathrm{charge}}(r) = \int \frac{d^{3}q}{(2\pi)^{3}} \:
  F_{Z,N}(q) \: \exp\left(-i \vec{q} \cdot \vec{r} \right) \: .
\end{equation}
It is used to calculate the root-mean-square charge radius
\begin{equation}
  \label{eq:r_c}
  r_{c} = \left[ \frac{1}{Z} \int d^{3}r \: r^{2} \: n_{\mathrm{charge}}(r)
    \right]^{1/2} \: , 
\end{equation}
of the nucleus. The form factor (\ref{eq:F_ZN})
at zero momentum has the values $F_{Z,N}(0) = Z$ and decreases with $q$
until it reaches a first zero at $q_{0}$. This value is used to obtain
the diffraction radius
\begin{equation}
  \label{eq:r_d}
  r_{d} = \frac{4.493409458}{q_{0}} \: ,
\end{equation}
where the numerical factor is the first zero of the Bessel function
$j_{1}(x)=-j_{0}^{\prime}(x)$. Further increasing $q$ leads to the first
minimum of $F_{Z,N}(q)$ at $q_{m}$ that is used to calculate
the surface thickness
\begin{equation}
  \sigma(Z,N) = \frac{1}{q_{m}}
  \left\{ 2 \ln
  \left[\frac{ZF_{\mathrm{box}}(q_{m})}{F_{Z,N}(q_{m})}\right]
  \right\}^{1/2} \: , 
\end{equation}
with the form factor
$F_{\mathrm {box}}(q) = 3 j_{1}(qr_{d})/(qr_{d})$ 
of a box of radius $r_{d}$,
c.f.\ Ref.\ \cite{Reinhard:1986qq}, and
Eq.\ (18a) of Ref.\ \cite{Rufa:1988zz} with corrections.
\begin{table}[t]
  \caption{Coefficients \cite{Simon:1980hu,Reinhard:1986qq}
    in the parameterisation (\ref{eq:F_pn}) of the nucleon
    form factors $F_{\eta}(q)$ ($\eta=\pi,\nu$).}
  \label{tab:F_pn}
  \begin{tabular}{@{}lcccc@{}}
    \toprule
    k & $A_{\pi k}$ & $B_{\pi k}$ & $A_{\nu k}$ & $B_{\nu k}$\\
      &          & [fm${}^{2}$] & & [fm${}^{2}$]\\
    \midrule
    1 &  $0.312$ & $0.16667$  &  $1$ & $0.04833$ \\
    2 &  $1.312$ & $0.06658$  & $-1$ & $0.05833$ \\
    3 & $-0.709$ & $0.02269$  &  $0$ &  $0$ \\
    4 &  $0.085$ & $0.006485$ &  $0$ &  $0$ \\
    \bottomrule
  \end{tabular}
\end{table}
The root-mean-square radius of a single-neutron state is defined as
\begin{equation}
  r_{\nu} = \left[  \int d^{3}r \: r^{2} \: n_{\nu k_{\nu}m_{\nu}}(r) \right]^{1/2} \: ,
\end{equation}
similar as the charge radius (\ref{eq:r_c}).
The spin-orbit splitting is given by the difference
\begin{equation}
  \Delta E_{so}(\eta n_{\eta}l_{\eta}) = E_{\eta k_{\eta}^{-}}-E_{\eta k_{\eta}^{+}} \: ,
\end{equation}
of the single-nucleon energies of the levels with
identical principal quantum number $n_{\eta}$, and
$k_{\eta}^{\pm} = (n_{\eta},\kappa_{\eta}^{\pm})$,
where
$\kappa_{\eta}^{\pm}=(j_{\eta}^{\pm}-l_{\eta})(2j_{\eta}^{\pm}+1)$
for $j_{\eta}^{\pm}=l_{\eta}\pm 1/2$.
The constraint isoscalar monopole giant resonance energy $E_{\mathrm{mono}}$
is obtained from a constraint calculation of the nucleus as described
in Ref.\ \cite{Agrawal:2005ix}. A contribution
\begin{equation}
  V_{\lambda} = -\lambda r^{2} \: ,
\end{equation}
with parameter $\lambda$,
is added to the vector potential (\ref{eq:V}) of the nucleons
and the coupled equations for the fields are solved self-consistently
for given $\lambda$.  Then the root-mean-square radius
\begin{equation}
  \langle r^{2} \rangle_{\lambda} = \int d^{3}r \: r^{2} \:
  n^{(v)}_{\lambda}(\vec{r}) \: , 
\end{equation}
of the total vector density $n^{(v)}_{\lambda}$ of the obtained
configuration is calculated.
The constraint isoscalar monopole giant resonance energy
\begin{equation}
  E_{\mathrm{mono}} = \sqrt{\frac{m_{1}}{m_{-1}}} \: ,
\end{equation}
can be found from the ratio of moments 
\begin{equation}
  m_{1} = 2 \frac{\hbar^{2}}{M_{\mathrm{nuc}}}
  \langle r^{2} \rangle_{\lambda=0} \: ,
\end{equation}
and,
\begin{equation}
  m_{-1} = \frac{1}{2} \left.
  \frac{d}{d\lambda} \langle r^{2} \rangle_{\lambda}  \right|_{\lambda=0} \: .
\end{equation} 
In the actual calculation, a five-point formula is used with
$\lambda_{n}=n \times 0.125$~keV and $n=-2,-1,0,1,2$
to calculate the derivative.

\subsubsection{Selection of nuclei and experimental data}

The description of nuclei with the relativistic EDF presented here can
only be applied to nuclei with spherical symmetry. Also pairing effects
are not taken into account. Thus a comparison of theory and experiment
can be reasonably performed only for magic or semi-magic nuclei. This limits
the amount of data that can be included in the fit of the model parameters.
However, this restriction is not too severe since the essential features
of strongly interacting systems are sufficiently well constrained by the
selection of the fitted data
in this approach. In order to fix the isospin dependence of the
effective interaction well, nuclei with large $|N-Z|$ should be included,
preferably on both sides of the valley of stability.
The nuclei actually considered can be read off from 
Tables \ref{tab:B_ZN}, \ref{tab:radii+}, and \ref{tab:monopole}.
There are 18 (semi)magic nuclei and 2 with
one valence neutron above a doubly magic core.
Not all data from the list in Subsection \ref{sec:obs}
are always available for all nuclei. 

Binding energies $B(Z,N)$
of nuclei are obtained from the Atomic Mass Evaluation 2020
\cite{Wang:2021xhn} by multiplying the binding energy per nucleon given there
with the mass number $A=Z+N$. These data
are tabulated in Table \ref{tab:B_ZN}.
Experimental data for the charge radius, diffraction radius, surface thicknesses
and neutron radii can be found in Table \ref{tab:radii+}.
The charge radii and quantities related to the charge form factor
are extracted from
\cite{DeVries:1987atn,Fricke:1995zz,Nadjakov:1994ozl,Angeli:2013epw}
using the Fourier-Bessel parameterisation if available.
The experimental valence neutron radii are taken from
\cite{Kalantar-Nayestanaki:1988mpc,Platchkov:1988}.
Experimental spin-orbit splittings are given in Table \ref{tab:sos}.
They are derived from the nuclear level schemes collected
at the National Nucleat Data Center \cite{ENSDF,Typel:2005ba}.
Experimental constraint isoscalar monopole giant resonance energies
\cite{Youngblood:1999zza}
are listed in Table \ref{tab:monopole} for selected nuclei.
The choice of the constraint isoscalar monopole giant resonance energy
as an observable in the fit greatly helps to obtain a reasonable
incompressibility coefficient
$K$ of nuclear matter, a quantity that
is often set to a predefined value since it is not well constrained
by other quantities of nuclei.

\subsection{Objective function and uncertainties}
\label{sec:uncert}

In order to determine optimal parameter values of the relativistic EDF,
an appropriate fitting procedure has to be applied. This is usually done
by minimizing an objective function that quantifies the deviation of the
theoretical predictions for certain observables from experimental data.
A crucial question is the size of the uncertainties that are used for the
different observables, e.g., in a $\chi^{2}$ minimization. In the
parameter fit of an EDF it is generally not reasonable to use the
experimental uncertainties because they are for most quantities
much smaller than the precision that can be achieved by an EDF. Instead,
often some 'reasonable' values are chosen that may, however, not reflect
the 'true' uncertainties that are expected for such an EDF.
Thus, in the present work a fitting scheme is employed that
also attempts to determine the uncertainties, c.f.\ Ref.\ \cite{Typel:2020ozc}.

In the present approach to fit the model parameters,
$N_{\mathrm{obs}}=7$ types of observables are considered
as specified in Section \ref{sec:obs} with
$N_{i}^{\mathrm{(obs)}}$ data per observable ($i=1,\dots,N_{\mathrm{obs}}$).
Thus an objective function,
\begin{equation}
  \label{eq:chi2}
  \chi^{2}(\{p_{k}\})
  =  \sum_{i=1}^{N_{\mathrm{obs}}}  \chi_{i}^{2}(\{p_{k}\}) \: ,
\end{equation}
is defined that depends on $N_{\mathrm{par}}$ parameters $p_{k}$
($k=1,\dots,N_{\mathrm{par}}$) with $N_{\mathrm{par}}=11$ $(9)$ for
models with (without) tensor couplings. The contributions
for each observable $\mathcal{O}_{i}$ to
(\ref{eq:chi2}) are given by the standard expression
\begin{equation}
  \label{eq:chi2i}
  \chi_{i}^{2}(\{p_{k}\})
  =  \sum_{n=1}^{N_{i}^{\mathrm{(obs)}}}
  \left[ \frac{\mathcal{O}_{i}^{\mathrm{(model)}}(n,\{p_{k}\})
      - \mathcal{O}_{i}^{\mathrm{(exp)}}(n)}{\Delta \mathcal{O}_{i}} \right]^{2}
  \: ,
\end{equation}
where the $n=1,\dots,N_{i}^{\mathrm{obs}}$
data of an observable $\mathcal{O}_{i}$ depend on the parameters $\{p_{k}\}$.
The total number of data is given by
\begin{equation}
  N_{\mathrm{data}} = \sum_{i=1}^{N_{\mathrm{obs}}} N_{i}^{\mathrm{(obs)}} \: ,
\end{equation}
and thus the number of degrees of freedom
is $N_{\mathrm{dof}} = N_{\mathrm{data}} - N_{\mathrm{par}}$. 

The uncertainties for each observable $\mathcal{O}_{i}$
are denoted by $\Delta \mathcal{O}_{i}$. Since binding energies of stable nuclei
are much more precisely known than
those of exotic nuclei, their uncertainties vary over
a large range. Thus the uncertainty of the observable $\mathcal{O}_{i} = B(Z,N)$
in the function (\ref{eq:chi2i}) is assumed to compose of two contributions as,
\begin{equation}
  \Delta \mathcal{O}_{i} =
  \left[ \left(\Delta \mathcal{O}_{i}^{\mathrm{(exp)}}\right)^{2}
    + \left(\Delta \mathcal{O}_{i}^{\mathrm{(fit)}}\right)^{2} \right]^{1/2} \: ,
\end{equation}
with an experimental uncertainty $\Delta \mathcal{O}_{i}^{\mathrm{(exp)}}$
and a fit uncertainty $\Delta \mathcal{O}_{i}^{\mathrm{(fit)}}$.
For all other observables, the experimental uncertainty is neglected in the
fits of this work. A reasonable estimate of the fit uncertainties will be
obtained if,
\begin{equation}
  \chi^{2}(\{p_{k}\})/N_{\mathrm{dof}} = 1  \: ,
\end{equation}
and,
\begin{equation}
  \frac{\chi_{i}^{2}(\{p_{k}\})}{N_{i}^{\mathrm{(obs)}}} =
  \frac{\chi^{2}(\{p_{k}\})}{N_{\mathrm{data}}} \: ,
\end{equation}
i.e., each observable contributes on the average equally
to the total $\chi^{2}$.
Theses conditions allow to determine the fit errors
$\Delta \mathcal{O}_{i}^{\mathrm{(fit)}}$ during the fit by a rescaling.
Thus the general fit procedure proceeds in two
steps: variation of the
model parameter to minimize (\ref{eq:chi2}) until a certain convergence
is reached, and a rescaling of the fit uncertainties.
These two steps are executed repeatedly until the fit uncertainties
and model parameters have converged to the desired precision.
To find the optimal values of the model parameters
in their space of dimension $N_{\mathrm{par}}$, a simplex method
\cite{NumRec} was employed as in Ref.\ \cite{Typel:2020ozc}. It only relies
on values of the $\chi^{2}$ function and does not need derivatives.
Once the minimum of the total $\chi^{2}$ function
is found, the solution can be represented as a vector
$\vec{p}^{\rm min}=(p_{1}^{\rm min}, \dots ,  p_{N_{\rm par}}^{\rm min})$
in the space of parameters.
A true (local) minimum of $\chi^{2}$ is reached when
the matrix of second derivatives, 
\begin{equation}
  \label{eq:Mij}
  \mathcal{M}_{ij} = \frac{1}{2}
  \left. \frac{\partial^{2} \chi^{2}}{\partial p_{i} \partial p_{j}}
  \right|_{\vec{p}_{\rm min}} \: ,
\end{equation}
has only positive eigenvalues.
If this is not the case, the eigenvectors with negative eigenvalues
indicate directions for a further reduction of $\chi^{2}$. This information
is used every now and then to accelerate the minimization. However, the
calculation of $\mathcal{M}_{ij}$ is expensive as it needs
$2N_{\mathrm{par}}^{2}+1$ function evaluations.
Finally, it is possible to determine the uncertainty
\begin{equation}
  \Delta \mathcal {O} =
  \sqrt{\overline{\left(\Delta \mathcal{O}\right)^{2}}} \: ,
\end{equation}
of any observable $\mathcal{O}$
and the correlation coefficient
\begin{equation}
  c_{\mathcal{O}_{1}\mathcal{O}_{2}} =
  \frac{\overline{\Delta \mathcal{O}_{1} \Delta \mathcal{O}_{2}}}{\sqrt{
      \overline{\left(\Delta \mathcal{O}_{1}\right)^{2}} \:
      \overline{\left(\Delta \mathcal{O}_{2}\right)^{2}}}} \: ,
\end{equation}
between two observables $\mathcal{O}_{1}$ and $\mathcal{O}_{2}$
from
\begin{equation}
  \overline{\Delta \mathcal{O}_{1} \Delta \mathcal{O}_{2}}
  = \sum_{ij} \frac{\partial \mathcal{O}_{1}}{\partial p_{i}}
  \left( \mathcal{M}^{-1}\right)_{ij}
  \frac{\partial \mathcal{O}_{2}}{\partial p_{j}} \: ,
\end{equation}
that contains the inverse of the matrix $\mathcal{M}$ and first derivatives
of the observables with respect to the parameters.

\subsection{Types of energy density functionals}

The main aim of the present work is the development of a new parameterisation
of a relativistic EDF with tensor couplings, denoted by 'DDT' in the
following, in addition to the usual
minimal couplings of the mesons to the nucleons.
In order to compare the main differences of this EDF
to a model without tensor interactions,
the results of the DD2 model \cite{Typel:2009sy} will be shown,
that uses the same form of the relativistic EDF.
Although a different fitting strategy with smaller
numbers of nuclei and observables was used to determine
the parameters of the DD2 model, a comparison of the results
with those of models with tensor couplings
will be instructive even though the differences will originate
not completely from these additional contributions in the EDF.

A particular issue is the effect of exchange contributions
that are neglected in the relativistic EDF, which is constructed 
in Hartree approximation from the Lagrangian density. This concerns
particularly the Coulomb interaction.
The source density of the electromagnetic field, that is
found by solving Eq.\ (\ref{eq:gamma}), includes the charge distribution
of all protons so that the Coulomb field in the vector potential
(\ref{eq:V}), which a proton feels in the mean-field created by all particles,
also contains a self-interaction contribution
that would not arise with the proper use
of an antisymmetrized many-body wave function.
The exchange term is sometimes added in form of the
Slater approximation \cite{Slater:1951}.
However, this is not done in this work. As an alternative,
a simple correction to
reduce the self-interaction term is to multiply the field $A_{0}$ in Eq.\
(\ref{eq:V}) by the factor $(Z-1)/Z$ in the calculation of nuclei,
as used in the DD2 parameterisation. Then the
vector potential $V_{\pi}(\vec{r})$ has the correct asymptotic form
for large radii
when a proton is separated from a nucleus. To study the effect of
this modification, a variation with this reduction factor,
DDTC, of the original DDT parameterisation is also developed.

A more detailed consideration of the contributions to the energy of a nucleus
reveals, however, that it is admissible to neglect the exchange part
in the Coulomb energy. In fact, there are two contributions that in total
compensate for the exchange term. On the one hand,
there is the Coulomb displacement energy
of analog states in mirror nuclei that, in the mean-field approximation, e.g.,
for the ${}^{41}$Sc - ${}^{41}$Ca pair is smaller by about 7\%.
This is known as the Nolen-Schiffer anomaly \cite{Nolen:1969ms}
which is solved by the taking into account the
charge symmetry breaking (CSB) contribution in the nucleon-nucleon
interaction, and
long-range correlation (LRC) effects \cite{Shlomo:1982}.
Ignoring the factor $(Z-1)/Z$
in the Coulomb direct term, which leads to including the
Coulomb self-interaction terms, and ignoring the Coulomb exchange term
corresponds to including the contributions of the CSB and LRC effects.
Thus only the
direct part of the Coulomb contribution to the energy should be included in
the calculation, see also subparagraph
IIA in Ref.\ \cite{Agrawal:2005ix}.

\section{Results and discussion}
\label{sec:res}

After performing the fit of the parameters for the two
models with tensor couplings,
various quantities for nuclei, nuclear matter and neutron stars
can be calculated and compared to experimental data. Before 
they are presented, in the following subsections, a discussion
of the model parameters and of the obtained
density dependence of the coupling functions
is of interest since they are essential to obtaining
a good quality fit.

The parameters of the three relativistic EDF models of the
present work are specified in Table \ref{tab:cpl}.
For the DDT and DDTC parameterisations,
uncertainties are given as determined with the method described
in Subsection \ref{sec:uncert}. For the DD2 model, these are not available
because the parameter set does not correspond to a minimum of $\chi^{2}$ with
the present set of observables and model uncertainties.
The most obvious differences are the large values of the tensor couplings
$\Gamma_{T\omega}$, and $\Gamma_{T\rho}$, and the small $\sigma$
meson masses $M_{\sigma}$
as compared to the value of the DD2 model
without tensor couplings. This is
not surprising since $\Gamma_{T\omega}$ and $M_{\sigma}$
directly affect the description
of the surface properties of nuclei and thus there is a strong correlation
between them. The DD2 model is also characterized by a rather small
saturation density.
The isoscalar couplings $\Gamma_{\sigma}^{(0)}$ and $\Gamma_{\omega}^{(0)}$
at the reference density of the DDT and DDTC models are considerably smaller
than those of the DD2 model. On the other hand, the couplings
$\Gamma_{\rho}^{(0)}$ and the coefficient $a_{\rho}$ are larger.
Particularly striking are the large coefficients $b_{\omega}$ and
$c_{\omega}$ with a small $d_{\omega}$ for the DDT model.
This will lead to an unusual density dependence
of the coupling $\Gamma_{\omega}$, see below.

\begin{table}[t]
  \caption{Parameters of the coupling functions,
    reference densities, and $\sigma$-meson masses
    for the different parameterisations of the relativistic EDF.
    Uncertainties of the quantities are given in brackets.}
  \label{tab:cpl}
  \begin{tabular}{@{}c|cc|cc|c@{}}
    \toprule
    parameter & \multicolumn{2}{|c|}{DDT} &
    \multicolumn{2}{|c|}{DDTC} & DD2 \\
    \midrule
    $\Gamma_{\sigma}^{(0)}$
    & $8.630588$ & $(0.289206)$ & $8.580686$ & $(0.846337)$ & $10.686681$ \\
    $a_{\sigma}$
    & $1.142684$ & $(0.009581)$ & $1.311104$ & $(0.048916)$ & $1.357630$ \\
    $b_{\sigma}$
    & $1.854752$ & $(0.627900)$ & $2.315517$ & $(2.062872)$ & $0.634442$ \\
    $c_{\sigma}$
    & $2.193260$ & $(0.741533)$ & $3.213878$ & $(2.826693)$ & $1.005358$ \\
    $d_{\sigma}$
    & $0.389847$ & $(0.065899)$ & $0.322051$ & $(0.146240)$ & $0.575810$ \\
    \midrule
    $\Gamma_{\omega}^{(0)}$
    & $11.14407$ & $(0.29709)$ & $11.09567$ & $(0.86419)$ & $13.342362$ \\
    $a_{\omega}$
    & $2.891633$ & $(1.032068)$ & $1.434242$ & $(0.204994)$ & $1.369718$ \\
    $b_{\omega}$
    & $1439.522$ & $(1290.251)$ & $9.292208$ & $(10.52430)$ & $0.496475$ \\
    $c_{\omega}$
    & $4164.427$ & $(5086.006)$ & $13.65208$ & $(17.02927)$ & $0.817753$ \\
    $d_{\omega}$
    & $0.008947$ & $(0.005460)$ & $0.156257$ & $(0.097456)$ & $0.638452$ \\
    \midrule
    $\Gamma_{\rho}^{(0)}$
    & $3.747188$ & $(0.070045)$ & $3.693565$ & $(0.086777)$ & $3.626940$ \\
    $a_{\rho}$
    & $0.684812$ & $(0.050891)$ & $0.675146$ & $(0.054904)$ & $0.518903$ \\
    \midrule
    $\Gamma_{T\omega}$
    & $3.200838$ & $(0.621091)$ & $3.489206$ & $(1.881621)$ & $0.000000$ \\
    $\Gamma_{T\rho}$
    & $18.67809$ & $(3.74586)$ & $20.50426$ & $(3.39598)$ & $0.000000$ \\
    \midrule
    $\varrho_{\mathrm{ref}}$ [fm$^{-3}$]
    & $0.15493496$ & ($0.00075982$) & $0.15304168$ & $(0.00094838)$ & $0.149065$ \\
    \midrule
    $M_{\sigma}$ [MeV]
    & $511.9820$ & $(6.5387)$ & $510.5184$ & $(19.2068)$ & $546.2125$ \\
    \bottomrule
  \end{tabular}
\end{table}

\begin{figure}[b]
\centering
\includegraphics[width=0.7\textwidth]{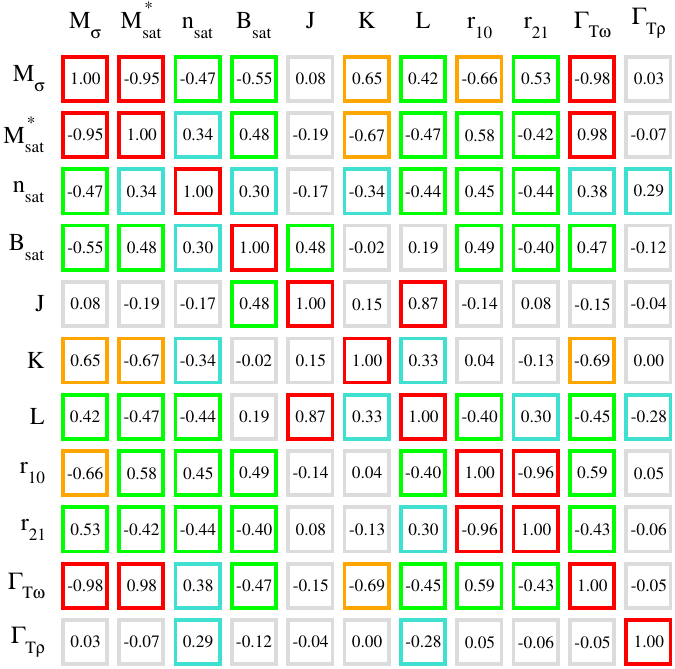}
\caption{Correlation matrix of the fit parameters in the DDT model.}
\label{fig:corr}
\end{figure}

The correlation coefficients of the fit parameters
in the DDT model are given in Figure \ref{fig:corr} where boxes in red color
indicate the largest absolute values
whereas grey boxes correspond to the smallest
moduli of the coefficients. As expected,
a first group with $M_{\sigma}$, $M_{\mathrm{sat}}^{\ast}$,
and $\Gamma_{T\omega}$ shows strong correlations since they all affect
the description of the nuclear surface. In contrast, $\Gamma_{T\rho}$
is hardly correlated to any other parameter of the fit. The other two
parameters related to the isovector part of the interaction,
$J$, and $L$, however, show a strong correlation. Apart from this
dependence, $J$ seems to be practically uncorrelated to all other
fit parametes. Another pair of parameters,
$r_{10}$, and $r_{21}$, determines the density dependence of the
isoscalar couplings. They are also strongly correlated. A somewhat lesser
strength of correlation is found for the incompressibility coefficient $K$,
and $r_{10}$ with the first group of parameters.

\begin{figure}[t]
\centering
\includegraphics[width=0.3\textwidth]{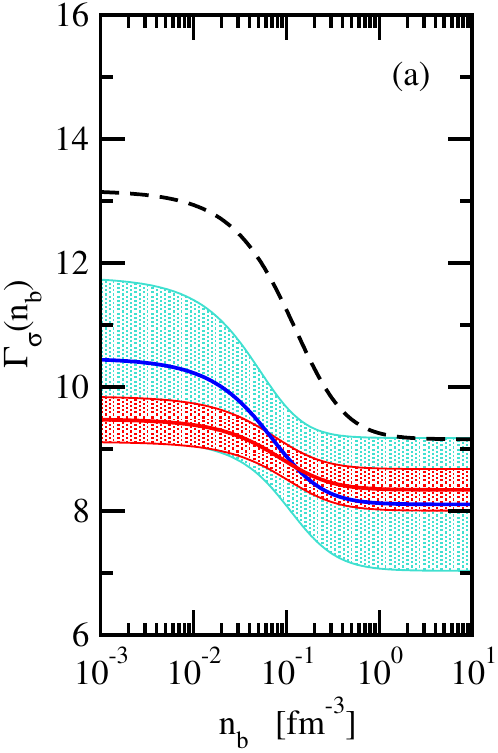}
\includegraphics[width=0.3\textwidth]{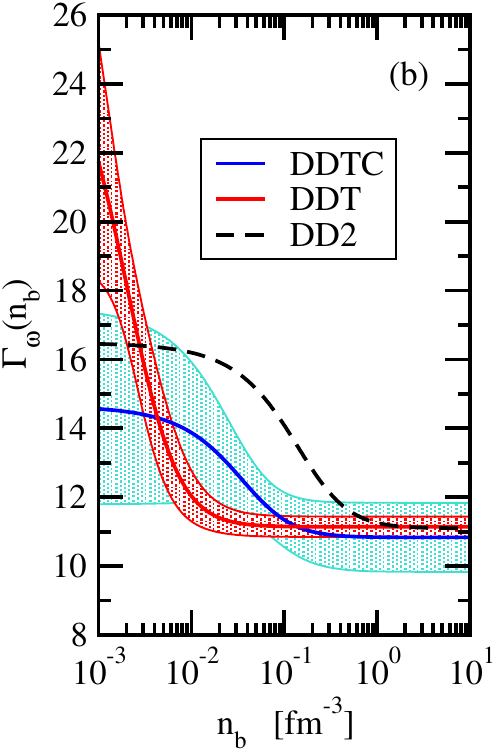}
\includegraphics[width=0.3\textwidth]{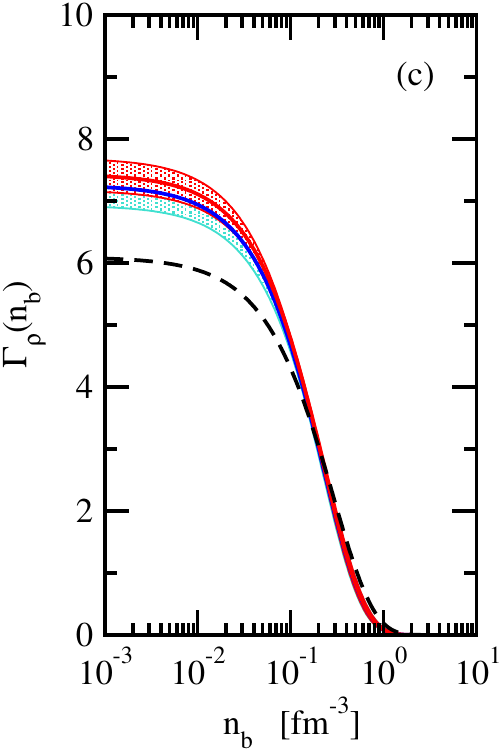}
\caption{Coupling functions $\Gamma_{i}(n_{b})$ of the
  (a) $\sigma$ meson, (b) $\omega$ meson, and (c) $\rho$ meson
  for the different parameterisations of the relativistic EDF
  with uncertainty bands for DDTC and DDT.}
\label{fig:Gamma}
\end{figure}

The coupling functions $\Gamma_{i}(n_{b})$ generally decrease
with increasing baryon density $n_{b}$, see Figure \ref{fig:Gamma}.
They approach a finite value in the high-density limit by construction, and
go to zero for $\Gamma_{\rho}$, in particular. The isoscalar couplings for the
models with tensor couplings (DDT, and DDTC)
are significantly smaller than those of the model without tensor couplings
(DD2). This can also be observed for the absolute values
of the couplings at the reference density in Table \ref{tab:cpl}.
The strong rise of the $\omega$ meson coupling
in the DDT model towards very small densities is particularly noteworthy.
This behavior is a consequence of the large values of the parameters
$b_{\omega}$ and $c_{\omega}$ in combination with a small $d_{\omega}$.
The isoscalar couplings, $\Gamma_{\omega}$ and $\Gamma_{\sigma}$,
of the DDTC parameterisation have a much larger uncertainty as compared to those
of the DDT model, but their shape is very similar to the corresponding
couplings of the DD2 model.
The density dependence of the
$\rho$ meson couplings is very similar for all models with slightly
higher values for the models with tensor couplings.

\subsection{Nuclei}
\label{sec:nuclei}

Before analyzing the predictions of individual properties
of finite nuclei, it is worthwhile to look at the fit uncertainties of the
observables, see Table \ref{tab:DO},
and their contributions
to the $\chi^{2}$ function that are collected in Table
\ref{tab:chi2} for the three parameterisations. The uncertainties of the
energy quantities, i.e., the binding energy, the spin-orbit splitting,
and the constraint isocalar monopole giant resonance energy, are all
on a sub-MeV level, whereas the uncertainties of
the size quantities, i.e., radii and skin thickness,
are typically in the order of a few hundredths of a fm with the
error in the root-mean-square (rms)
valence neutron radius beeing the smallest. These results give
an impression on the actual quality a description of nuclei with
a relativistic EDF can achieve and how reasonable values have to be chosen
in other attempts to find parameters of such models from fits
to finite nuclei.

\begin{table}[t]
  \caption{Fit uncertainties $\Delta \mathcal{O}_{i}^{(\mathrm{fit})}$
    of the observables $\mathcal{O}_{i}$ for DDT.}
  \label{tab:DO}
  \begin{tabular}{@{}cccccccc@{}}
    \toprule
    observable $\mathcal{O}_{i}$
    & $B$ & $r_{c}$ & $r_{d}$ & $\sigma$ & $r_{n}$ & $\Delta E_{so}$ & $E_{\mathrm{mono}}$\\
    & [MeV] & [fm] & [fm] & [fm] & [fm] & [MeV] & [MeV] \\
    \midrule
    fit uncertainty $\Delta \mathcal{O}_{i}^{\mathrm{(fit)}}$
    & 0.619311 & 0.013364 & 0.017155 & 0.026851 & 0.008249 & 0.240832 & 0.430714  \\
    \bottomrule
  \end{tabular}
\end{table}

\begin{table}[b]
  \caption{Contributions $\chi_{i}^{2}(\{p_{k}\})$ of the different
    observables and total $\chi^{2}(\{p_{k}\})$ for the different
    parameterisations of the relativistic EDF.
    The bottom line gives the sum of the quantities
    in the same column above.}
  \label{tab:chi2}
  \begin{tabular}{@{}ccrrr@{}}
    \toprule
    observable $i$ & $N_{i}^{\mathrm{(obs)}}$ & DDT & DDTC & DD2 \\
    \midrule
    $B(Z,N)$         & 16 & 12.48 & 31.79 & 100.02 \\
    $r_{c}$           & 8 & 6.24 & 12.61 & 16.06 \\
    $r_{d}$           & 5 & 3.90 & 6.30 & 26.27 \\
    $\sigma$         & 5 & 3.90 & 5.08 & 2.24 \\
    $r_{n}$           & 2 & 1.56 & 4.01 & 578.17 \\
    $\Delta E_{so}$   & 10 & 7.80 & 7.47 & 40.66 \\
    $E_{\mathrm{mono}}$ &  4 & 3.12 & 2.86 & 3.61 \\
    \midrule
    & 50 & 39.00 & 70.11 & 767.01 \\
    \bottomrule
  \end{tabular}
\end{table}

A comparison of the total $\chi^{2}$, see Table \ref{tab:chi2},
and the contributions for the three models reveals interesting differences.
First of all, it is evident that the DDT model gives the best description
of the data with the lowest total $\chi^{2}$. By construction, its value
is given by the number of degrees of freedom $N_{\mathrm{dof}}=39$.
The other parameterisations, in particular the one without tensor couplings,
are considerably worse. In case of the DDTC parameterisation
this is mostly caused by the larger errors in the
binding energies and the spin-orbit splittings.
The values of $\chi^{2}$ of the DD2 model is completely
dominated by the difference of the valence neutron radii from the experimental
values. Even neglecting their contribution to $\chi^{2}$, the descripton
of experimental data with the DD2 model is almost always much worse than
that of the DDT and DDTC models. All observables
that are determined strongly by the surface description of a nucleus, viz.,
the radii, the skin thickness, and the spin-orbit splitting,
are on the average better described in the EDFs with tensor couplings.
Only the surface thickness $\sigma$ is slightly better described
by the DD2 model.

\begin{table}[t]
  \caption{Binding energies $B(Z,N)$ of nuclei in MeV from
    experiment \cite{Wang:2021xhn} in comparison to theoretical values
    from different parameterisations of the relativistic EDF.
    Uncertainties of the quantities are given in brackets.
    The bottom line gives the root-mean-square deviation
    of the theoretical predictions from the experiment in MeV.}
  \label{tab:B_ZN}
  \begin{tabular}{@{}c|c|cc|cc|c@{}}
    \toprule
    nucleus     &  experiment & \multicolumn{2}{|c|}{DDT} &
    \multicolumn{2}{|c|}{DDTC} & DD2 \\
    \midrule
    ${}^{16}$O   &  127.619 &  128.193 & (0.262) &  128.469 & (0.285) &  127.947 \\
    ${}^{24}$O   &  168.952 &  168.368 & (0.327) &  167.890 & (0.440) &  169.130 \\
    ${}^{28}$O   &  167.664 &  167.888 & (0.701) &  167.672 & (0.839) &  173.005 \\
    ${}^{34}$Si  &  283.464 &  284.471 & (0.300) &  283.732 & (0.300) &  284.044 \\
    ${}^{34}$Ca  &  243.882 &  244.060 & (0.234) &  245.259 & (0.259) &  246.245 \\
    ${}^{40}$Ca  &  342.052 &  342.161 & (0.362) &  342.366 & (0.444) &  342.203 \\
    ${}^{48}$Ca  &  416.001 &  416.545 & (0.224) &  416.225 & (0.265) &  414.929 \\
    ${}^{48}$Ni  &  347.328 &  346.740 & (0.290) &  348.870 & (0.347) &  350.092 \\
    ${}^{56}$Ni  &  483.996 &  483.435 & (0.380) &  483.523 & (0.412) &  482.346 \\
    ${}^{68}$Ni  &  590.408 &  589.675 & (0.374) &  588.546 & (0.367) &  590.595 \\
    ${}^{78}$Ni  &  642.564 &  642.398 & (0.470) &  642.460 & (0.464) &  642.048 \\
    ${}^{90}$Zr  &  783.897 &  782.939 & (0.321) &  782.332 & (0.376) &  780.727 \\
    ${}^{100}$Sn &  825.163 &  825.984 & (0.490) &  825.978 & (0.551) &  826.231 \\
    ${}^{132}$Sn & 1102.843 & 1103.173 & (0.394) & 1103.256 & (0.456) & 1103.388 \\
    ${}^{140}$Ce & 1172.683 & 1172.895 & (0.423) & 1173.098 & (0.625) & 1172.814 \\
    ${}^{208}$Pb & 1636.430 & 1636.354 & (0.531) & 1636.477 & (0.530) & 1635.922 \\
    \midrule
    & & 0.562 & & 0.919 & & 1.907 \\
    \bottomrule
  \end{tabular}
\end{table}

\begin{figure}[b]
\centering
\includegraphics[width=0.75\textwidth]{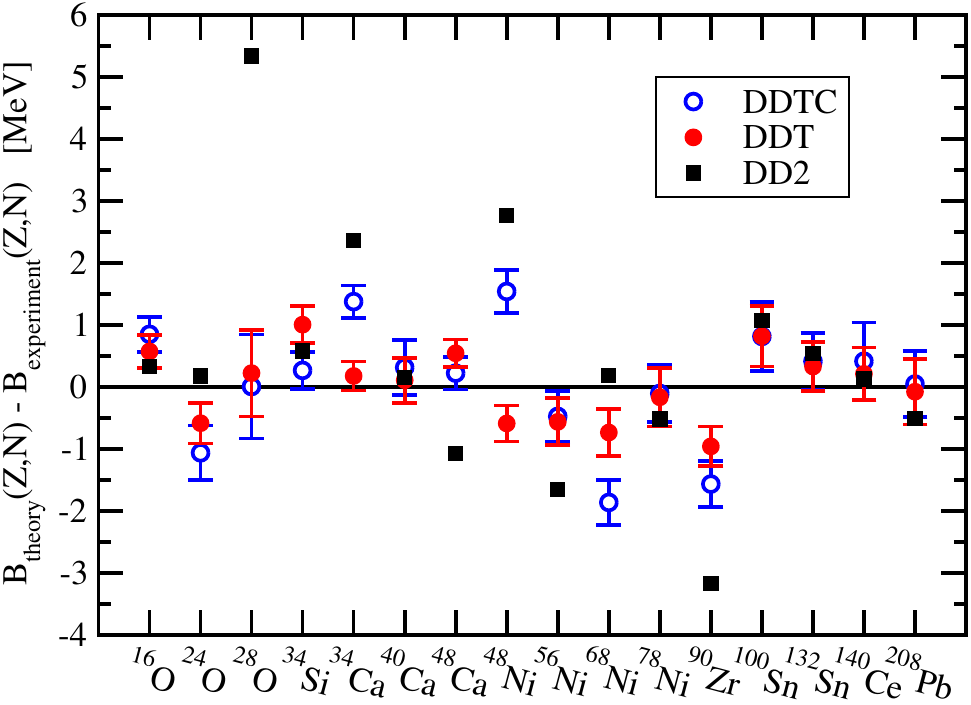}
\caption{Binding energy differences
  $B_{\mathrm{theory}}(Z,N)-B_{\mathrm{experiment}}(Z,N)$
  of nuclei used in the parameter fits with theoretical
  uncertainties indicated by the error bars.}
\label{fig:dB}
\end{figure}

The first quantity of major interest is the binding energy
$B(Z,N)$ of a nucleus that is predicted by a given parameterisation
of a relativistic EDF.
The results of the binding energy calculations
of the three models of the present work
are listed in Table \ref{tab:B_ZN}, for nuclei considered in the fit,
in comparison to experimental values from the Atomic Mass Evaluation
2020 \cite{Wang:2021xhn}. In order to have a better impression
of the quality of the models, the differences between the experimental
and theoretical data are depicted in Figure \ref{fig:dB}. It is seen
that the models with tensor coupling show less fluctuations and deviate
on the average less strongly from the experimental data than the models without
tensor couplings. This is also reflected in the root-mean-square deviation
of the theoretical values from the experimental data in the bottom
line of Table \ref{tab:B_ZN}. The trend of these numbers agrees with the
variation of the binding energy contribution to the total $\chi^{2}$
shown in Table \ref{tab:chi2}.

Also interesting observations for individual sets of nuclei
can be made. In constrast to the DD2 parameterisation,
the binding energy of ${}^{28}$O is smaller than that of ${}^{24}$O
in the DDT and DDTC parameterisations, a result that is difficult
to achieve in most other relativistic EDFs.
Similar patterns can be seen in Figure \ref{fig:dB} when
comparing different parameterisations, e.g., for the heaviest nuclei.
There are also striking effects for
the set of Ni nuclei with different trends when comparing the models.
It is seen in Table \ref{tab:B_ZN}
that the calculated binding energies of ${}^{48}$Ni
obtained for the DDT model without the $(Z-1)/Z$ factor
is significantly closer to the experimental value than that of the
DDTC and DD2 models with the $(Z-1)/Z$ factor.

\begin{table}[t]
  \caption{Experimental data of charge radii, diffraction radii,
    surface thicknesses
    \cite{DeVries:1987atn,Fricke:1995zz,Nadjakov:1994ozl,Angeli:2013epw}
    and valence neutron radii
    \cite{Kalantar-Nayestanaki:1988mpc,Platchkov:1988} in fm
    of nuclei used in the fit,
    in comparison to theoretical values
    from different parameterisations of the relativistic EDF.
    Uncertainties of the quantities
    are given in brackets.}
  \label{tab:radii+}
  \begin{tabular}{@{}c|c|c|cc|cc|c@{}}
    \toprule
    & nucleus     &  experiment & \multicolumn{2}{|c|}{DDT} &
    \multicolumn{2}{|c|}{DDTC} & DD2 \\
    \midrule
    charge radius $r_{c}$
    & ${}^{16}$O    & 2.699 & 2.705 & (0.005) & 2.689 & (0.011) & 2.691  \\
    & ${}^{40}$Ca   & 3.478 & 3.453 & (0.003) & 3.440 & (0.004) & 3.463  \\
    & ${}^{48}$Ca   & 3.477 & 3.489 & (0.005) & 3.482 & (0.007) & 3.478  \\
    & ${}^{68}$Ni   & 3.887 & 3.871 & (0.004) & 3.864 & (0.004) & 3.887  \\
    & ${}^{90}$Zr   & 4.269 & 4.269 & (0.004) & 4.264 & (0.004) & 4.282  \\
    & ${}^{132}$Sn  & 4.709 & 4.707 & (0.005) & 4.708 & (0.005) & 4.732  \\
    & ${}^{140}$Ce  & 4.877 & 4.886 & (0.005) & 4.888 & (0.005) & 4.897  \\
    & ${}^{208}$Pb  & 5.501 & 5.504 & (0.006) & 5.507 & (0.008) & 5.540  \\
    \midrule
    diffraction radius $r_{d}$
    & ${}^{16}$O    & 2.764 & 2.767 & (0.006) & 2.756 & (0.011) & 2.707  \\
    & ${}^{40}$Ca   & 3.849 & 3.832 & (0.004) & 3.825 & (0.005) & 3.816  \\
    & ${}^{48}$Ca   & 3.963 & 3.973 & (0.005) & 3.971 & (0.010) & 3.953  \\
    & ${}^{90}$Zr   & 5.040 & 5.015 & (0.005) & 5.015 & (0.006) & 5.019  \\
    & ${}^{208}$Pb  & 6.776 & 6.787 & (0.008) & 6.799 & (0.009) & 6.829  \\
    \midrule
    surface thickness $\sigma$
    & ${}^{16}$O    & 0.851 & 0.828 & (0.005) & 0.810 & (0.008) & 0.837  \\
    & ${}^{40}$Ca   & 0.968 & 0.964 & (0.005) & 0.946 & (0.009) & 0.968  \\
    & ${}^{48}$Ca   & 0.890 & 0.909 & (0.005) & 0.899 & (0.006) & 0.897  \\
    & ${}^{90}$Zr   & 0.957 & 1.001 & (0.005) & 0.993 & (0.007) & 0.994  \\
    & ${}^{208}$Pb  & 0.919 & 0.914 & (0.004) & 0.906 & (0.006) & 0.912  \\
    \midrule
    valence neutron radius $r_{n}$
    & ${}^{17}$O    & 3.360 & 3.356 & (0.008) & 3.377 & (0.008) & 3.514  \\
    & ${}^{41}$Ca   & 3.990 & 3.999 & (0.006) & 3.990 & (0.011) & 4.116  \\    
    \bottomrule
  \end{tabular}
\end{table}

\begin{figure}[t]
\centering
\includegraphics[width=0.28\textwidth]{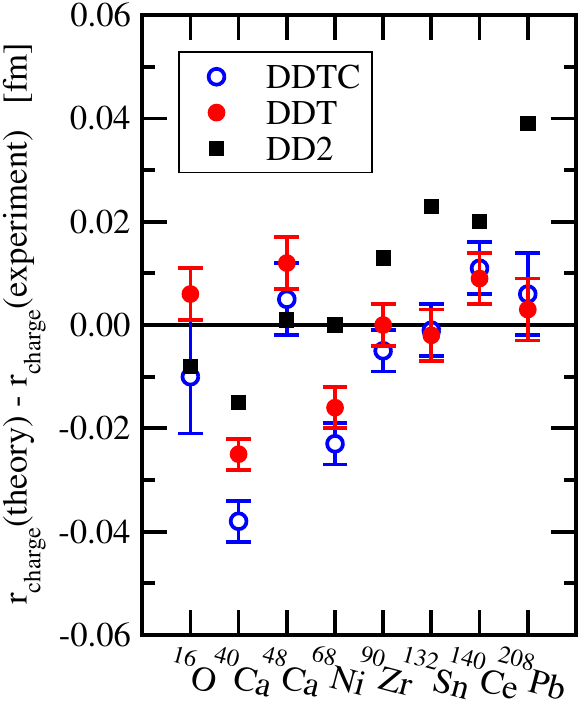}
\hspace{1ex}
\includegraphics[width=0.28\textwidth]{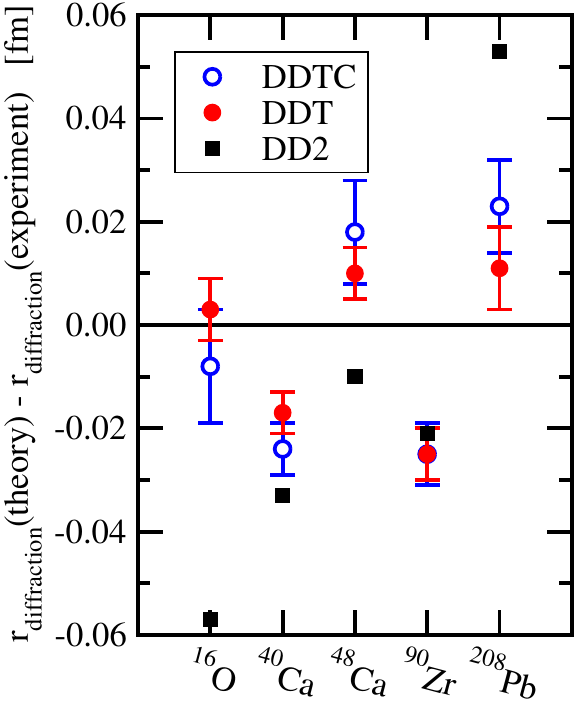}
\hspace{1ex}
\includegraphics[width=0.28\textwidth]{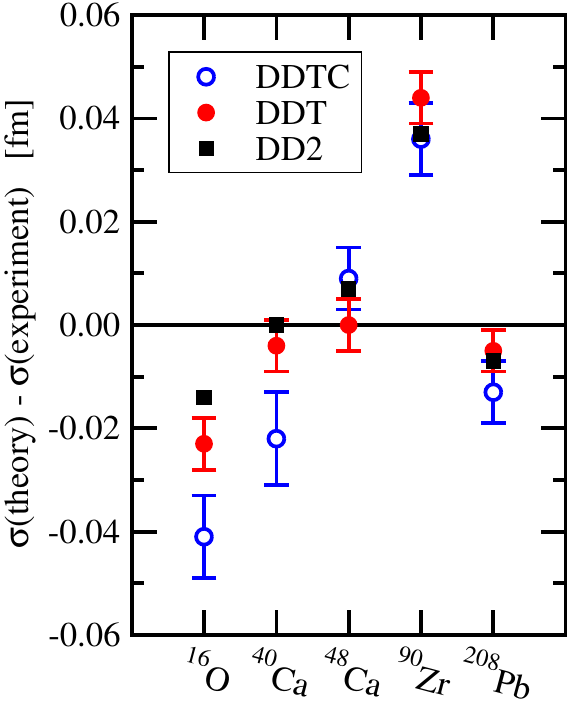}
\caption{Differences between theoretical and experimental
  charge radii $r_{c}$,
  diffraction radii $r_{d}$ and surface thicknesses $\sigma$,
  respetively,
  of nuclei used in the fit of the model parameters.
  Theoretical uncertainties are indicated by the error bars.}
\label{fig:diff_rcrdsurf}
\end{figure}

Theoretical values of the quantities related to the 
charge form factor are given in Table \ref{tab:radii+} in comparison
to experimental data for the nuclei in the fit. The theoretical
uncertainties are generally in the order of 0.005~fm.
Differences between theory and experiment
are depicted in Figure \ref{fig:diff_rcrdsurf} with three panels
for the charge radius, the diffraction radius, and the surface
thickness. There is a good agreement with typical differences
between theory and experiment of
less than 0.02~fm, and a clear trend in the deviations
is not seen. Only the surface thickness of ${}^{90}$Zr is overestimated
by more than 0.04~fm. The values of the valence neutron radii of ${}^{17}$O
and ${}^{41}$Ca can also be found in Table \ref{tab:radii+}. The
deviations between theory and experiment
are slightly larger than those of the
quantities related to the charge form factor, but the agreement with
the experimental data is still satisfactory, given the large
theoretical uncertainties.

The differences between experimental and theoretical
values of the spin-orbit splittings can be explored by studying
Table \ref{tab:sos}. Considering the theoretical uncertainties of
somewhat less than 0.1~MeV, there is a fair agreement of these
data. The difference between the experimental
spin-orbit splittings of the $\nu 0 p$ and $\pi 0p$ levels
in ${}^{16}$O ($-0.140$~MeV) is not reproduced by the theoretical models
as their differences are positive, $0.182$~MeV and
$0.170$~MeV for the DDT and DDTC models, respectively.
The spin-orbit splitting of the $\nu 0 p$ levels is
larger than that of the $\pi 0 p$ levels in contrast
to the experiment, in particular for the DD2 parameterisation.
The spin-orbits splittings of the
$\nu 1 p$ levels in ${}^{48}$Ca are underestimated in the theoretical
models in comparison to the experimental value.
Larger differences between theory and experiment
are also observed for the $\pi 1d$ levels in
${}^{132}$Sn, and ${}^{208}$Pb.
In general, the results of the DD2 model show a larger
deviation from the experiment as compared to the models with tensor
couplings.

The contributions with tensor couplings in the EDF
modify the strength of the spin-orbit interaction as compared to
models without these couplings. Thus a modification in the single-particle
level structure can be expected. This could also affect the appearance of
shell gaps. As an example, the semi-magic proton shell gaps in ${}^{140}$Ce
and ${}^{218}$U at $Z=58$ and $Z=92$, respectively, are examined here.
It has been observed that many relativistic EDFs predict large
spurious shell gaps for these nuclei, see, e.g., Refs.\
\cite{Geng:2019qah,Wei:2020kfb,Zhao:2022xhq}.
In the DD2 model the $\pi 0 g_{7/2}- \pi 1 d_{5/2}$ splitting in ${}^{140}$Ce
amounts to $3.054$~MeV and the $\pi 0 h_{9/2} - \pi 1 f_{7/2}$ splitting
in ${}^{218}$U is $3.248$~MeV. In fact, reduced values of
$2.288$~MeV ($2.543$~MeV) and $2.665$~MeV ($2.999$~MeV)
are found for the $Z=58$ and $Z=92$ proton shell gaps in the DDT (DDTC) model.

\begin{table}[b]
  \caption{Experimental data of spin-orbit splittings
    \cite{ENSDF,Typel:2005ba} in MeV
    of single-particle levels in nuclei used in the fit
    in comparison to theoretical values
    from different parameterisations of the relativistic EDF.
    Uncertainties of the quantities
    are given in brackets.}
  \label{tab:sos}
  \begin{tabular}{@{}c|c|c|cc|cc|c@{}}
    \toprule
    nucleus     & level &  experiment & \multicolumn{2}{|c|}{DDT} &
    \multicolumn{2}{|c|}{DDTC} & DD2 \\
    \midrule
    ${}^{16}$O   & $\nu 0 p$ & 6.180 & 6.127 & (0.095) & 6.231 & (0.214) & 6.867 \\
                & $\pi 0 p$ & 6.320 & 5.945 & (0.094) & 6.061 & (0.209) & 6.819 \\
    \midrule
    ${}^{48}$Ca  & $\nu 1 p$ & 2.020 & 1.841 & (0.068) & 1.849 & (0.075) & 1.510 \\
    \midrule
    ${}^{90}$Zr  & $\pi 1 p$ & 1.510 & 1.598 & (0.049) & 1.649 & (0.121) & 1.701 \\
    \midrule
    ${}^{132}$Sn & $\pi 1 d$ & 1.480 & 1.749 & (0.080) & 1.825 & (0.202) & 2.014 \\
                & $\pi 0 g$ & 6.140 & 6.254 & (0.097) & 6.083 & (0.169) & 6.720 \\
    \midrule
    ${}^{140}$Ce & $\nu 2 p$ & 0.475 & 0.560 & (0.068) & 0.622 & (0.049) & 0.517 \\
    \midrule
    ${}^{208}$Pb & $\nu 2 p$ & 0.898 & 1.066 & (0.065) & 0.993 & (0.037) & 0.930 \\
                & $\pi 1 d$ & 1.330 & 1.680 & (0.037) & 1.732 & (0.064) & 1.868 \\
                & $\pi 0 h$ & 5.550 & 5.712 & (0.088) & 5.527 & (0.085) & 6.203 \\    
    \bottomrule
  \end{tabular}
\end{table}

The energies of the isoscalar monopole giant
resonance are crucial, in the fit of the model parameters,
to obtain
a reasonable value of the incompressibility coefficient $K$ of
nuclear matter. Experimental data and theoretical values are
given in Table \ref{tab:monopole} with theoretical uncertainties
in the order of 0.2~MeV. In general, there is a good reproduction
of the experimental values by theory, by the models with and without
tensor couplings. Only the prediction for the
resonance energy of ${}^{208}$Pb is on the lower side as compared to the
experiment.

\begin{table}[t]
  \caption{Experimental data of
    constraint isoscalar monopole giant resonance
    energies \cite{Youngblood:1999zza}
    in MeV of nuclei used in the fit
    in comparison to theoretical values
    from different parameterisations of the relativistic EDF.
    Uncertainties of the quantities
    are given in brackets.}
  \label{tab:monopole}
  \begin{tabular}{@{}c|c|cc|cc|c@{}}
    \toprule
    nucleus     &  experiment & \multicolumn{2}{|c|}{DDT} &
    \multicolumn{2}{|c|}{DDTC} & DD2 \\
    \midrule
    ${}^{90}$Zr   & $17.810$ & $17.597$ & (0.214) & 17.714 & (0.214) & 17.584 \\
    \midrule
    ${}^{116}$Sn  & $15.900$ & $16.112$ & (0.193) & 16.189 & (0.196) & 16.085 \\
    \midrule
    ${}^{144}$Sm  & $15.250$ & $15.158$ & (0.212) & 15.220 & (0.318) & 15.205 \\
    \midrule
    ${}^{208}$Pb  & $14.180$ & $13.489$ & (0.188) & 13.519 & (0.232) & 13.417 \\
    \bottomrule
  \end{tabular}
\end{table}

\begin{table}[b]
  \caption{Theoretical neutron skin thickness $\Delta r_{\mathrm{skin}}$
    in fm of selected nuclei used in the fit
    from different parameterisations of the relativistic EDF.
    Uncertainties of the quantities
    are given in brackets.}
  \label{tab:nskin}
  \begin{tabular}{@{}c|rc|rc|r@{}}
    \toprule
    nucleus     & \multicolumn{2}{|c|}{DDT} &
    \multicolumn{2}{|c|}{DDTC} & DD2 \\
    \midrule
    ${}^{16}$O   & $-0.031$ & (0.001) & $-0.029$ & (0.001) & $-0.029$  \\
    ${}^{24}$O   & $ 0.450$ & (0.014) & $0.480$ & (0.018) & $ 0.595$  \\
    ${}^{28}$O   & $ 0.622$ & (0.019) & $0.650$ & (0.042) & $ 0.769$  \\
    ${}^{34}$Si  & $ 0.181$ & (0.007) & $0.183$ & (0.007) & $ 0.203$  \\
    ${}^{34}$Ca  & $-0.299$ & (0.007) & $-0.302$ & (0.008) & $-0.323$  \\
    ${}^{40}$Ca  & $-0.053$ & (0.001) & $-0.052$ & (0.001) & $-0.053$  \\
    ${}^{48}$Ca  & $ 0.125$ & (0.008) & $0.127$ & (0.008) & $ 0.191$  \\
    ${}^{48}$Ni  & $-0.251$ & (0.007) & $-0.251$ & (0.006) & $-0.316$  \\
    ${}^{56}$Ni  & $-0.057$ & (0.001) & $-0.057$ & (0.001) & $-0.054$  \\
    ${}^{68}$Ni  & $ 0.166$ & (0.008) & $0.167$ & (0.009 & $ 0.200$  \\
    ${}^{78}$Ni  & $ 0.251$ & (0.012) & $0.249$ & (0.013) & $ 0.329$  \\
    ${}^{90}$Zr  & $ 0.040$ & (0.007) & $0.039$ & (0.007) & $ 0.083$  \\
    ${}^{100}$Sn & $-0.083$ & (0.001) & $-0.083$ & (0.001) & $-0.082$  \\
    ${}^{132}$Sn & $ 0.191$ & (0.011) & $0.189$ & (0.013) & $ 0.261$  \\
    ${}^{140}$Ce & $ 0.086$ & (0.009) & $0.083$ & (0.010) & $ 0.150$  \\
    ${}^{208}$Pb & $ 0.134$ & (0.010) & $0.130$ & (0.013) & $ 0.198$  \\    
    \bottomrule
  \end{tabular}
\end{table}

In the determination of the model
parameters, several exotic nuclei on both sides of the valley of
stability were included in order to better constrain the isovector
part of the effective in-medium interaction. In contrast to
most earlier relativistic EDFs with density dependent couplings,
three independent isovector
parameters are considered in the parameterisations
that include tensor couplings. It is interesting to study, how this
affects the predictions of the neutron skin thickness of nuclei,
$\Delta r_{\mathrm{skin}} = r_{\nu}-r_{\pi}$,
that is obtained by the difference
between the neutron and proton rms radii. The theoretical values
for the nuclei in the fit are collected in Table \ref{tab:nskin}.
Nuclei with $N=Z$ exhibit a small negative value of
$\Delta r_{\mathrm{skin}}$ that becomes even more negative
for neutron deficient nuclei like ${}^{34}$Ca and ${}^{48}$Ni.
On the other hand, neutron-rich nuclei can develop a
substantial neutron skin. This is most clearly seen for
${}^{24}$O, ${}^{28}$O, ${}^{78}$Ni, and ${}^{132}$Sn.
The theoretical uncertainties, however, become much larger
for the most exotic nuclei as compared to stable nuclei.
The neutron skin thicknesses of more exotic nuclei are generally
larger for the DD2 models as compared to the DDT and DDTC models. 
Also stable nuclei such as ${}^{48}$Ca and ${}^{208}$Pb
show a sizeable neutron skin thickness. It can be determined
experimentally by studying the parity-violation asymmetry in
high-energy electron scattering experiments.
A comparison of the theoretical predictions in Table
\ref{tab:nskin} with the experimental data of the
CREX experiment \cite{CREX:2022kgg} with
$\Delta r_{\mathrm{skin}}({}^{48}\mbox{Ca}) =
0.121\pm 0.025 \: \mbox{(exp.)} \pm 0.024 \: \mbox{(model)}$~fm
and those of the combined PREX experiments \cite{PREX:2021umo}
with $\Delta r_{\mathrm{skin}}({}^{208}\mbox{Pb}) = 0.283\pm 0.071$~fm
shows that the theory can nicely reproduce the size of the
neutron skin in ${}^{48}$Ca within the uncertainties
but it predicts a considerably smaller
neutron skin thickness in ${}^{208}$Pb, an issue that
is not solved so far. The neutron skin thickness can
also be determined using hadronic probes using precisely
known proton radii from electron scattering.
A former analysis of 1~GeV proton scattering
on ${}^{48}$Ca \cite{Shlomo:1979hbl} gave a neutron
skin thickness of $\Delta r_{\mathrm{skin}}=0.10 \pm 0.03$~fm only
slightly smaller but consistent within the errors with the CREX
result. A more recent study of neutron radii using proton scattering
\cite{Clark:2002se} gave ranges of
$0.056~\mbox{fm} < \Delta r_{\mathrm{skin}}({}^{48}\mbox{Ca}) < 0.102$~fm
and
$0.083~\mbox{fm} < \Delta r_{\mathrm{skin}}({}^{208}\mbox{Pb}) < 0.111$~fm
that point to rather small neutron thicknesses of these
nuclei.

The inclusion of pairs of mirror nuclei
in the fit also allows to better constrain the mirror displacement
energies. The binding energy differences of three pairs of
mirror nuclei are given in Table \ref{tab:dB_mirror}. There is a good
agreement between experiment and theory with theoretical results
of the DDT model that are somewhat larger than the experimental values.
The differences between experimental and theoretical values
are $0.374$,  $0.883$, and $1.132$~MeV for the pairs with mass number
$A=41$, $34$, and $48$ for the DDT model, respectively, thus overestimating the
binding energy differences. In contrast, they are systematically smaller
with $0.072$, $-1.055$, and $-1.318$~MeV for the DDTC model
and $-0.237$, $-1.738$, and $-3.836$~MeV for the DD2 model, respectively,
where the $Z(Z-1)$ factor in the Coulomb field was introduced.
In addition, 
the neutron-deficient nuclei were not considered in the fit of the parameters
in the latter parameterisation.
Thus the parameter set DDT gives the best description on the average.
The theoretical uncertainties increase with the isospin difference of the
pair of nuclei.
  
\begin{table}[t]
  \caption{Experimental binding energy
      differences in MeV of mirror nuclei in comparison
      to theoretical values
      from different parameterisations of the relativistic EDF.
      Uncertainties of the quantities
      are given in brackets.}
  \label{tab:dB_mirror}
  \begin{tabular}{@{}c|c|cc|cc|c@{}}
    \toprule
    pair of mirror nuclei & experiment &
    \multicolumn{2}{|c|}{DDT} & \multicolumn{2}{|c|}{DDTC} & DD2 \\
     \midrule
     ${}^{41}$Ca - ${}^{41}$Sc &  7.278 & 7.652 & (0.031)
     &  7.350 & (0.030) &  7.041 \\
     \midrule
     ${}^{34}$Si - ${}^{34}$Ca & 39.528 & 40.411 & (0.170)
     & 38.473 & (0.168) & 37.790 \\
     \midrule
     ${}^{48}$Ca - ${}^{48}$Ni & 68.673 & 69.805 & (0.272)
     & 67.355 & (0.244) & 64.837 \\
    \bottomrule
  \end{tabular}
\end{table}


\subsection{Nuclear matter}

\begin{figure}[b]
\centering
\includegraphics[width=0.75\textwidth]{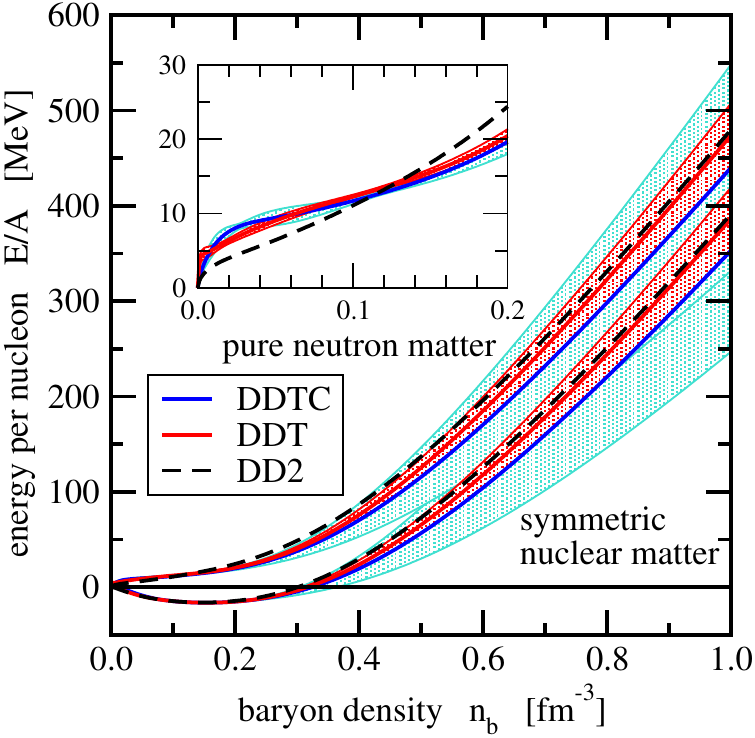}
\caption{Equation of state of symmetric nuclear matter
  and pure neutron matter, respectively, with
  theoretical uncertainty bands for the DDT and DDTC models.
  The inset shows the low-energy equation of state
  of pure neutron matter.}
\label{fig:eos}
\end{figure}

The energy per nucleon $E/A$, Eq.\  (\ref{eq:EA}),
in homogeneneous nuclear matter is a
function of the baryon density $n_{b}=n_{\nu}+n_{\pi}$ and the
isospin asymmetry $\alpha = (n_{\nu}-n_{\pi})/n_{b}$.
For $\alpha=0$, the energy per nucleon of symmetric nuclear matter
$E_{0}=E/A(n_{b},0)$ is obtained. It is depicted in Figure
\ref{fig:eos} for the 
parameterisations DDT, DDTC, and DD2.
The minimum of $E_{0}$ for these
equations of state defines the saturation density
$n_{\mathrm{sat}}$, see Table \ref{tab:nucmat}.
The region of negative slope, i.e., negative pressure, signals an
instability of homogeneous matter that is related to the
occurrence of the liquid-gas phase transition.
At high densities, the predictions for the
energy per nucleon become less certain as the widening
of the uncertainty bands of the DDT and,
significantly larger, DDTC models shows.
The energy per nucleon
of pure neutron matter, i.e., $E/A$ for $\alpha=1$,
is also shown in Figure \ref{fig:eos}.
It increases monotonically with the
baryon density.

The curves of the DDT and DD2
parameterisations in Fig.\ \ref{fig:eos} look very similar
at first sight. But there are differences in detail, in particular
at lower densities.
The curves of the parameterisation DDTC follow along the
lower limit of the
uncertainty bands of the DDT model, hence signaling a softer equation of state.
The neutron matter equation of state of the DDT model
shows a peculiar feature at the lowest densities. This can be seen
in the inset of Figure \ref{fig:eos}. The energy per nucleon does not
approach zero as smoothly as in the DD2 and DDTC models.
This is related to the
specific density dependence of the $\omega$-meson coupling $\Gamma_{\omega}$,
see panel (b) of Figure \ref{fig:Gamma}. The increase of the coupling
at very small densities causes an additional repulsion, which also applies
for symmetric nuclear matter at very low densities (not shown in the inset).

\begin{figure}[t]
\centering
\includegraphics[width=0.75\textwidth]{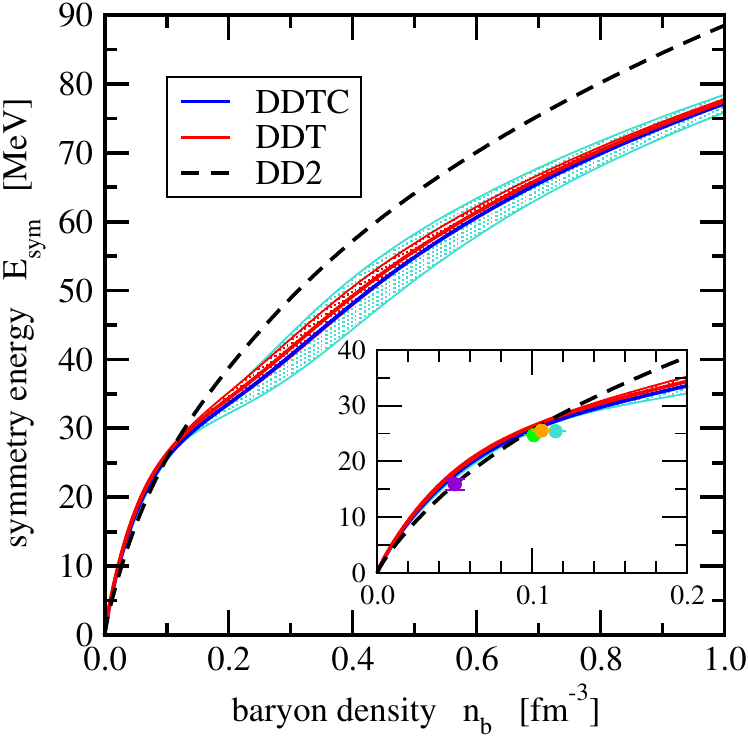}
\caption{Symmetry energy $E_{\mathrm{sym}}$ as function
  of the baryon density $n_{b}$ with
  theoretical uncertainty bands for the DDT and DDTC models.
  The inset shows the low density part with a comparison to
  experimental data, see text for details.}
\label{fig:esym}
\end{figure}

Assuming a quadratic dependence of $E/A$ on
$\alpha$ for given baryon density $n_{b}$, the symmetry
energy (\ref{eq:Esym}) can be obtained as a difference
$E_{\mathrm{sym}}(n_{b})=E/A(n_{b},1)-E/A(n_{b},0)$. This quantity
is shown in Figure \ref{fig:esym} as function of $n_{b}$.
Particularly noteworthy is the change of the slope
above saturation density,
for the DDT and DDTC models,
with a widening of the uncertainty bands.
At densities below saturation, symmetry energy is
well constrained by experiments and ab-initio calculations
of pure neutron matter, see, e.g., Refs.\
\cite{Chatziioannou:2024tjq,Tews:2024owl} and references therein.
At densities above saturation,
the predictions for the equation of state of nuclear matter
are less precise as indicated by the increasing uncertainty
bands.

The $E_{\mathrm{sym}}$ curves of the DDT, DDTC, and DD2 model
cross at a density below $n_{\mathrm{sat}}$.
It has been observed before that fitting model parameters of EDFs
to properties of nuclei effectively
determines the symmetry energy around this subsaturation density,
see, e.g., Ref.\ \cite{Lynch:2021xkq}. A comparison of the low-density
symmetry energy of the DDT, DDTC, and DD2 models with experimental data 
is depicted in the inset of Figure \ref{fig:esym}. These data come from
the determination of the dipole polarizability of ${}^{208}$Pb (violet),
the analysis of EDFs (green, light blue), and isobaric analog states (orange),
see Ref.\ \cite{Tsang:2023vhh} and references therein.
The symmetry energy of the
parameterisations DDT and DDTC lie slightly above these points, corresponding
to the small slope parameter $L$, whereas the DD2 curve
matches the data somewhat better.

Characteristic nuclear matter parameters
are given in Table \ref{tab:nucmat} for the
models DDT, DDTC, and DD2.
Most remarkable are the increased effective mass
and the reduced incompressibility coefficient
of the models with tensor couplings
(DDT and DDTC) in comparison to that without these couplings (DD2).
Most existing relativistic models show values of $K$ that are rather high
when compared with non-relativistic EDFs, cf.\ Refs.\
\cite{Dutra:2014qga,Dutra:2012mb}.
The tensor models are an exception and they
are also characterized by rather large saturation densities.
The higher-order coefficients $Q$ and $K_{\mathrm{sym}}$ show considerable
differences and are not very well determined by the parameter fits as they have
large uncertainties.
The smaller values of $K$ and $Q$ of the DDTC model
indicate a softer equation of state as compared to the DDT model as seen already
in Figure \ref{fig:eos}.
The DDT and DDTC models also have symmetry energy slope
parameters that are smaller than the value of $L$ of the DD2 model. This
difference correlates with the smaller neutron skin thicknesses for exotic
nuclei as discussed in Subsection \ref{sec:nuclei}.
  
\begin{table}[t]
  \caption{Characteristic nuclear matter parameters
    for different parameterisations of the relativistic EDF.
    Uncertainties of the quantities
    are given in brackets.}
  \label{tab:nucmat}
  \begin{tabular}{@{}cc|cc|cc|c@{}}
    \toprule
    quantity  &  unit & \multicolumn{2}{|c|}{DDT} &
    \multicolumn{2}{|c|}{DDTC} & DD2 \\
    \midrule
    $n_{\mathrm{sat}}$
    & fm${}^{-3}$ & $0.15493$ & $(0.00076)$ & $0.15304$ & $(0.00095)$ & $0.14907$ \\
    $M_{\mathrm{sat}}^{\ast}$ & $M_{\mathrm{nuc}}$
    & $0.65729$ & $(0.01315)$ & $0.66305$ & $(0.03803)$ & $0.56252$ \\
    $B_{\mathrm{sat}}$ 
    & MeV & $16.226$ & $(0.025)$ & $16.226$ & $(0.040)$ & $16.022$ \\
    $K$
    & MeV & $229.20$ & $(7.99)$ & $215.05$ & $(7.71)$ & $242.72$ \\
    $Q$
    & MeV & $88.565$ & $(230.061)$ & $-291.86$ & $(763.87)$ & $168.77$ \\
    $J$
    & MeV & $31.209$ & $(0.498)$ & $30.379$ & $(0.568)$ & $31.670$ \\
    $L$
    & MeV & $34.236$ & $(4.332)$ & $33.317$ & $(7.847)$ & $55.033$ \\
    $K_{\mathrm{sym}}$
    & MeV & $-69.538$ & $(15.454)$ & $-75.562$ & $(26.416)$ & $-93.226$ \\
    $K_{\tau,v}$
    & MeV & $-288.18$ & $(54.09)$ & $-230.25$ & $(131.21)$ & $-461.69$ \\
    \bottomrule
  \end{tabular}
\end{table}


\subsection{Neutron stars}

\begin{figure}[t]
\centering
\includegraphics[width=0.8\textwidth]{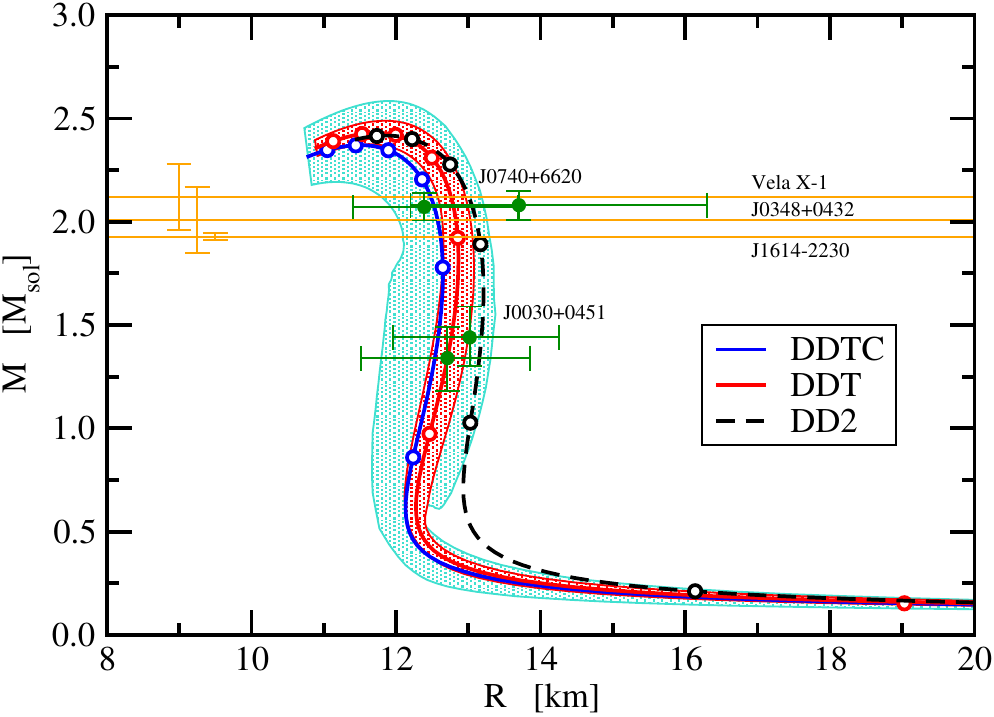}
\caption{Mass-radius relation of neutron stars
  for the parameterisations DDTC (blue), DDT (red), with
  uncertainty bands, and DD2 (black)
  of the relativistic EDF. Open circles indicate neutron stars
  where the central density is an integer multiple
  $n_{\mathrm{central}}=k n_{\mathrm{sat}}$ of the saturation density
  starting with $k=1$ at the lowest masses.
  (The point with $k=1$ of the DDTC parameterisation is at
    a radius $R=21.62$~fm outside the figure.)
  Yellow lines give the masses of heavy neutron stars
  \cite{Falanga:2015mra,Antoniadis:2013pzd,Fonseca:2016tux}
  with errors indicated on the left. Masses and radii of neutron stars from
  NICER observations
  \cite{Miller:2019cac,Riley:2019yda,Miller:2021qha,Riley:2021pdl}
  are shown in green color.}
\label{fig:nstar}
\end{figure}

The properties of a 
non-rotating, spherically symmetric neutron star, are obtained
by solving the first-order
Tolman-Oppenheimer-Volkoff equation
\cite{Tolman:1939jz,Oppenheimer:1939ne}
\begin{equation}
  \label{eq:TOV}
  \frac{dP}{dr} = -G \: \frac{M(r)\varepsilon(r)}{r^{2}}
  \left[ 1 + \frac{P(r)}{\varepsilon(r)}\right]
  \left[ 1 + \frac{4\pi r^{3}P(r)}{M(r)}\right]
  \left[ 1 - \frac{2 G M(r)}{r} \right]^{-1} \: , 
\end{equation}
with the mass
\begin{equation}
  M(r) = 4\pi \int_{0}^{r} dr^{\prime} \: (r^{\prime})^{2} \:
  \varepsilon(r^{\prime}) \: ,
\end{equation}
inside the volume of
radius $r$. The factors in square brackets are
the general relativistic corrections to the classical
expression for the change of the pressure with the radius.
Assuming a particular density
$n_{\mathrm{central}}$
in the center of the star, the first order differential equation
(\ref{eq:TOV}) can be solved by an outward integration until
the pressure $P(r)$ becomes sufficiently small. This defines the
stellar radius $R$ and the total mass $M(R)$. The essential
ingredient in the calculation is the equation of state
of charge-neutral matter in $\beta$ equilibrium. It is obtained
by adding the leptonic contributions (electrons and muons) to the
nuclear equation of state and determining the isospin asymmetry
$\alpha$ for given baryon density $n_{b}$, by requiring vanishing
lepton chemical potentials $\mu_{e}+\mu_{q}$, and
$\mu_{\mu}+\mu_{q}$, with charge chemical potential
$\mu_{q}=\mu_{\pi}-\mu_{\nu}$.
At densities below nuclear saturation,
an appropriate crust equation of state has to be used as the
matter is no longer homogeneous. Here, the
crust description of the generalized relativistic density functional
with the DD2 parameterisation of the couplings \cite{Typel:2009sy}
is used where the masses of the nuclei in the crust are taken from
the Duflo-Zuker model DZ31 \cite{Duflo:1995ep},
if not available from the atomic mass
evaluation AME2020 \cite{Wang:2021xhn}.

The mass-radius relations of neutron stars
for the different parameterisations
are depicted in Figure \ref{fig:nstar}. Open circles on
the curves denote the properties of neutron stars
with central densities that are an integer multiple
of the saturation density $n_{\mathrm{sat}}$.
The curves show the well-known inverse
S-shaped rise of the mass with increasing central density
and the corresponding variation of the radius. Ordinary neutron
stars with masses of 1.4 times the solar mass $M_{\mathrm{sol}}$
have central densities that are typically between twice
and three times the saturation density. At certain central
densities $n_{\mathrm{central}}^{(\mathrm{max})}$, the most massive
neutron star configurations are obtained.
They have central densities of five to six times
the saturation density. Neutron stars with
higher central density are not stable any more and this last
branch of the curve will not be populated. The maximum masses
of the shown models are always above the largest
masses of observed neutron stars
\cite{Falanga:2015mra,Antoniadis:2013pzd,Fonseca:2016tux}
since their equation of state
is sufficiently stiff. Masses and radii are avalaible from
NICER observations
\cite{Miller:2019cac,Riley:2019yda,Miller:2021qha,Riley:2021pdl},
but these data have large uncertainties. All models are consistent
with these observations.
Characteristic properties of neutron stars
of the models in this work are collected in Table \ref{tab:nstar}.
The properties of the maximum mass stars are very similar in all models
but the radii of the DD2 model are systematically larger that those
of the parameterisation DDT and DDTC. The latter model
predicts the lowest maximum mass, a higher central density at this mass,
and smaller radii as compared to the other models,
corresponding to the softer equation of state.

\begin{table}[t]
  \caption{Properties of neutron stars for the different
    parameterisations of the relativistic EDF.}
  \label{tab:nstar}
  \begin{tabular}{@{}ccccc@{}}
    \toprule
    quantity & unit & DDT & DDTC & DD2  \\
    \midrule
    $R_{1.4}$ & km & 12.744 & 12.566 & 13.172 \\
    $M_{\mathrm{max}}$ & $M_{\mathrm{sol}}$ & 2.430 & 2.371 & 2.417 \\
    $R_{\mathrm{max}}$ & km & 11.739 & 11.504 & 11.869 \\
    $n^{(\mathrm{max})}_{\mathrm{central}}$ & fm$^{-3}$
    & 0.85785 & 0.89663 & 0.85114 \\    
    \bottomrule
  \end{tabular}
\end{table}

\section{Summary and Conclusions}
\label{sec:concl}

In this work, an extension of a relativistic EDF is
presented that includes tensor couplings between the Lorentz vector mesons,
$\omega$ and $\rho$, and nucleons in addition to the standard minimal
meson-nucleon couplings. The latter are assumed to depend on the baryon
density. The field equation are derived in the mean-field approximation
from a covariant Lagrangian density
and solved self-consistently for nuclei and nuclear matter.

The parameters of this phenomenological approach are found from a fit
to properties of 20 selected nuclei that comprise binding energies, charge and
diffraction radii, surface thicknesses, spin-orbit splittings,
valence neutron radii,
and constraint isoscalar monopole giant resonance energies.
The model uncertainties
of these quantities are determined self-consistently during
the fitting procedure.
The model predictions are compared to experimental data and given with
their theoretical uncertainties, which were, to our knowledge,
not specified in earlier parameterisations of relativistic EDFs.
Two new sets of parameters, DDT and DDTC, are developed that differ
in the treatment of the Coulomb interaction.

The inclusion of tensor couplings in the relativistic EDF
leads to a considerable improvement
in the description of nuclear properties, in particular their binding energies.
Quantities related to the nuclear charge form factor show a similar quality
as those of an earlier parameterisation (DD2) without tensor couplings.
The binding energy differences of mirror nuclei are also closer to the
experimental data than in the DD2 model, that did not consider exotic,
neutron deficient nuclei in the determination of the parameters.
Predictions for the neutron skin thickess of bound and unstable nuclei are made.

The equations of state for symmetric nuclear matter and pure neutron matter
are calculated with the new parameters. They look similar as those of the
DD2 model but show some specific differences. These are most
apparent by comparing characteristic nuclear matter parameters.
The new parameterisations
exhibit larger saturation densities and effective
nucleon masses, and smaller incompressibility coefficients and symmetry
energy slope parameters.

The description of charge-neutral nuclear matter in $\beta$ equilibrium
with the extended EDFs allows to calculate properties of neutron stars and,
in particular, their mass-radius relation. The results are similar as
for the DD2 model but with slightly smaller radii. Due to the stiffness
of the EOS at high densities, the maximum masses 
are predicted to lie above the data of observed heavy neutron stars.
The obtained mass-radius relations are consistent with results from
NICER measurements that still have substantial uncertainties, in particular
for the radii.

EDFs with the new parameterisations can be applied
in further calculations to explore their quality in the description of finite nuclei.
The study of deformed nuclei is of particular interest as it allows to investigate
additional observables, e.g., higher multipole moments of the density distributions,
or the variation of energies of single-particle states
in isotopic and isotonic chains of nuclei.
The tensor couplings could affect the  position of the levels
in the spectrum and give rise to the
appearance of new shell gaps or their suppression. Effects of
pairing have to be included in the model too when open-shell nuclei are considered. 
However, additional work with proper numerical methods
is required to include deformation and pairing in the calculations.
This is beyond the scope of the present work.
The new models can also be used to prepare EOS tables
of stellar matter at finite temperature for larger ranges in isospin
asymmetry and density, in order to be used in simulations of core-collapse
supernovae and neutron-star mergers \cite{Oertel:2016bki}.
For this, the calculations have to
be extended by including additional degrees of freedom, e.g., light clusters
at subsaturation densities \cite{Typel:2009sy,Typel:2018wmm},
or hyperons at high densities, see, e.g., Ref.\ \cite{Stone:2019blq}.

\section*{Acknowledgments}

S. Typel acknowledges the support by the Cyclotron Institute,
Texas A\&M University.
S. Shlomo is partially supported by the US Department of Energy under Grant no. DE-FG03-93ER-40773.

\section*{Statements and Declarations}


Open Access funding enabled and organized by
Projekt DEAL.
The authors have no relevant financial or non-financial interests to disclose.
Data sets generated during the current study are available from the corresponding author on reasonable request.

\begin{appendices}
  
  \section{Transformation between parameters of nuclear matter and
    of coupling functions}
  \label{sec:appA}

  The parameters of the coupling functions (\ref{eq:Gamma_dd}) can be derived
  from the characteristic nuclear matter parameters.
  To simplify this conversion, the average nucleon mass
  $M_{\mathrm{nuc}}=(M_{\pi}+M_{\nu})/2$ is used instead of proton and
  neutron masses in the full calculation.
  However, this does not pose any problem.
  \begin{enumerate}
  \item The reference density
    $\varrho_{\mathrm{ref}}$ is chosen as the
    baryon density at saturation $n_{\mathrm{sat}}$ and defines the
    Fermi momentum
    \begin{equation}
      k_{\mathrm{sat}}^{\ast} =
      \left( \frac{3\pi^{2}}{2} n_{\mathrm{sat}}\right)^{1/3} \: .
    \end{equation}
  \item The Dirac effective mass at saturation $M_{\mathrm{sat}}^{\ast}$
    gives the effective chemical potential
    \begin{equation}
      \mu_{\mathrm{sat}}^{\ast} = \sqrt{(k_{\mathrm{sat}}^{\ast})^{2}
       + (M_{\mathrm{sat}}^{\ast})^{2}} \: ,
    \end{equation}
    and the scalar density at saturation
    \begin{equation}
      n_{\mathrm{sat}}^{(s)} = \frac{M_{\mathrm{sat}}^{\ast}}{\pi^{2}}
        \left[ k_{\mathrm{sat}}^{\ast} \mu_{\mathrm{sat}}^{\ast}
          - (M_{\mathrm{sat}}^{\ast})^{2}
          \ln \left( \frac{k_{\mathrm{sat}}^{\ast}
            +\mu_{\mathrm{sat}}^{\ast}}{M_{\mathrm{sat}}^{\ast}} \right) \right]
        \: .
    \end{equation}
    This determines the scalar coupling at saturation
    \begin{equation}
      C_{\sigma} =
      \frac{M_{\mathrm{nuc}}-M_{\mathrm{sat}}^{\ast}}{n_{\mathrm{sat}}^{(s)}} \: ,
    \end{equation}
    and the auxiliary quantity
    \begin{equation}
      f_{\mathrm{sat}} = 3 \left(
      \frac{n_{\mathrm{sat}}^{(s)}}{M_{\mathrm{sat}}^{\ast}}
       - \frac{n_{\mathrm{sat}}}{\mu_{\mathrm{sat}}^{\ast}} \right) \: ,
    \end{equation}
    that will be used below.
  \item The energy per nucleon at saturation $(E/A)_{\mathrm{sat}}$
    gives the energy density at saturation
    \begin{equation}
      \varepsilon_{\mathrm{sat}} = \left[ (E/A)_{\mathrm{sat}}
       + M_{\mathrm{nuc}}\right] n_{\mathrm{sat}} \: ,
    \end{equation}
    and the vector coupling at saturation
    \begin{equation}
      C_{\omega} = \frac{2}{n_{\mathrm{sat}}^{2}}
      \left[ \varepsilon_{\mathrm{sat}}
        - \frac{1}{4} \left( 3 \mu_{\mathrm{sat}}^{\ast}  n_{\mathrm{sat}}
        + M_{\mathrm{sat}}^{\ast} n_{\mathrm{sat}}^{(s)} \right)
        - \frac{1}{2} C_{\sigma} \left( n_{\mathrm{sat}}^{(s)} \right)^{2} \right]
       \: .
    \end{equation}
    Furthermore, the chemical potential
    \begin{equation}
      \mu_{\mathrm{sat}} = \frac{\varepsilon_{\mathrm{sat}}}{n_{\mathrm{sat}}}
       =  (E/A)_{\mathrm{sat}} + M_{\mathrm{sat}} \: ,
    \end{equation}
    and the rearrangement potential
    \begin{equation}
      V^{(R)} = \mu_{\mathrm{sat}} - \mu_{\mathrm{sat}}^{\ast}
       - C_{\omega} n_{\mathrm{sat}} \: ,
    \end{equation}
    are found.
  \item The ratio $r_{10}$ determines the derivatives
    \begin{equation}
      C_{\omega}^{\prime} = 2 r_{10} \frac{C_{\omega}}{n_{\mathrm{sat}}} \: ,
    \end{equation}
    and
    \begin{equation}
      C_{\sigma}^{\prime} = \frac{1}{\left( n_{\mathrm{sat}}^{(s)} \right)^{2}}
      \left( C_{\omega}^{\prime} n_{\mathrm{sat}}^{2}
      - 2 V^{(R)}\right) \: ,
    \end{equation}
    at saturation.
    Then, the second and third auxiliary variables
    \begin{equation}
      g_{\mathrm{sat}} =
      \frac{\frac{M_{\mathrm{sat}}^{\ast}}{\mu_{\mathrm{sat}}^{\ast}}
        - C_{\sigma}^{\prime}n_{\mathrm{sat}}^{(s)}f_{\mathrm{sat}}}{1+
      C_{\sigma}f_{\mathrm{sat}}} \: ,
    \end{equation}
    and,
    \begin{equation}
      h_{\mathrm{sat}} = - \frac{C_{\sigma}^{\prime}n_{\mathrm{sat}}^{(s)}
        + C_{\sigma} \frac{M_{\mathrm{sat}}^{\ast}}{\mu_{\mathrm{sat}}^{\ast}}}{1
      + C_{\sigma}f_{\mathrm{sat}}} \: ,
    \end{equation}
    can be calculated.
  \item The ratio
    $r_{21}$ is used to obtain the second derivative
    \begin{equation}
      C_{\omega}^{\prime\prime} = 2 r_{10} (r_{10}+r_{21})
      \frac{C_{\omega}}{n_{\mathrm{sat}}^{2}} \: ,
    \end{equation}
    for the $\omega$ meson at saturation.
  \item The incompressibility coefficient $K$ gives the
    second derivative,
    \begin{eqnarray}
      \nonumber C_{\sigma}^{\prime\prime} & = &
      \frac{2}{n_{\mathrm{sat}}\left( n_{\mathrm{sat}}^{(s)} \right)^{2}}
      \left[ \frac{1}{3} \left( \mu_{\mathrm{sat}}^{\ast}
        - M_{\mathrm{sat}}^{\ast} g_{\mathrm{sat}}\right)
        - \frac{K}{9}
        + \left( C_{\omega} + 2 C_{\omega}^{\prime} n_{\mathrm{sat}}
        + \frac{1}{2} C_{\omega}^{\prime\prime} n_{\mathrm{sat}}^{2}
        \right) n_{\mathrm{sat}}
        \right. \\ 
        & & \left.
        - \left( C_{\sigma} + C_{\sigma}^{\prime} n_{\mathrm{sat}}
        \right) n_{\mathrm{sat}}^{(s)} g_{\mathrm{sat}}
        - C_{\sigma}^{\prime} \left( n_{\mathrm{sat}}^{(s)} \right)^{2}
        \right] \: ,
    \end{eqnarray}
    for the $\sigma$ meson.
  \item The symmetry energy at saturation $J$ is used to calculate
    the coupling of the $\rho$ meson at saturation,
    \begin{equation}
      \Gamma_{\rho}^{(0)} = \sqrt{C_{\rho}} M_{\rho} \: ,
    \end{equation}
    where
    \begin{equation}
      C_{\rho} = \frac{2}{n_{\mathrm{sat}}}
      \left[ J - \frac{\left(k_{\mathrm{sat}}^{\ast}\right)^{2}}{6
          \mu_{\mathrm{sat}}^{\ast}}\right] \: .
    \end{equation}
  \item The slope parameter $L$ of the symmetry energy at saturation
    gives the derivative
    \begin{equation}
      C_{\rho}^{\prime} = \frac{2}{3n_{\mathrm{sat}}^{2}}
      \left\{ L - 3J - \frac{\left(k_{\mathrm{sat}}^{\ast}\right)^{2}}{3
        \mu_{\mathrm{sat}}^{\ast}}
      \left[ \frac{1}{2}
        \left( \frac{M_{\mathrm{sat}}^{\ast}}{\mu_{\mathrm{sat}}^{\ast}}
        \right)^{2} \left( 1 - 3 \frac{n_{\mathrm{sat}}}{M_{\mathrm{sat}}^{\ast}}
        h_{\mathrm{sat}}\right) - 1 \right]
      \right\} \: ,
    \end{equation}
    for the $\rho$ meson at saturation.
    Then the parameter
    \begin{equation}
      a_{\rho} = -\frac{1}{2} \frac{C_{\rho}^{\prime}}{C_{\rho}} n_{\mathrm{sat}} \: ,
    \end{equation}
    in the function $f_{\rho}(x)$ is found.
  \item The values of $C_{j}$ and their derivatives determine
    the couplings and first derivatives
    \begin{equation}
      \Gamma_{j}^{(0)} = \sqrt{C_{j}} M_{j} \: ,
      \qquad \mbox{and} \qquad
      \Gamma_{j}^{\prime} = \frac{1}{2}\Gamma_{j}^{(0)}
      \frac{C_{j}^{\prime}}{C_{j}} \: ,
    \end{equation}
    as well as the second derivatives
    \begin{equation}
      \Gamma_{j}^{\prime\prime} = \frac{1}{2} \Gamma_{j}^{(0)}
      \left[ \frac{C_{j}^{\prime\prime}}{C_{j}}
        - \frac{1}{2} \left( \frac{C_{j}^{\prime}}{C_{j}} \right)^{2}\right] \: ,
    \end{equation}
    for $j=\sigma$ and $\omega$.
    Then also the derivatives 
    \begin{equation}
      f_{j}^{\prime}(1) = \frac{\Gamma_{j}^{\prime}}{\Gamma_{j}^{(0)}} \: ,
      \qquad \mbox{and} \qquad
      f_{j}^{\prime\prime}(1) = \frac{\Gamma_{j}^{\prime\prime}}{\Gamma_{j}^{(0)}} \: ,
    \end{equation}
    of the functions $f_{j}(x)$ are known.
  \item Finally, the parameters of the rational functions $f_{j}(x)$
    for the isoscalar mesons can be found with help of the explicit
    expressions;
    \begin{eqnarray}
      f_{j}^{\prime}(x) & = & 2 a_{j}(b_{j}-c_{j})
      \frac{x+d_{j}}{[1+c_{j}(x+d_{j})^{2}]^{2}} \: ,
      \\
      f_{j}^{\prime\prime}(x) & = & 2 a_{j}(b_{j}-c_{j})
      \frac{1-3c(x+d_{j})^{2}}{[1+c_{j}(x+d_{j})^{2}]^{3}} \: ,
    \end{eqnarray}
    for the first and second derivatives of the functions $f_{j}(x)$. The ratio,
    \begin{equation}
      r_{21} = \frac{f_{\omega}^{\prime\prime}(1)}{f(1)}
       = \frac{d_{j}^{2}-(1+d_{j})^{2}}{[d_{j}^{2}+(1+d_{j})^{2}/3](1+d_{j})} \: ,
    \end{equation}
    is a function of $d_{j}$ only because of the constraint
    $1=3c_{j}d_{j}^{2}$
    that follows from $f_{j}^{\prime\prime}(0)=0$. Thus the value of $d_{j}$ can
    be obtained by a simple numerical search or iteration.
    Then $c_{j}=1/(3d_{j}^{2})$ is known
    as well and $b_{j}$ is calculated from
    \begin{equation}
      b_{j} = \frac{c_{j} z + r_{10}}{z-(1+d_{j})^{2} r_{10}} \: ,
    \end{equation}
    with the auxiliary quantity
    \begin{equation}
      z = \frac{2(1+d_{j})}{1+c_{j}(1+d_{j})^{2}} \: .
    \end{equation}
    Finally, the normalization condition $f_{j}(1)=1$ gives
    \begin{equation}
      a_{j} = \frac{1+c_{j}(1+d_{j})^{2}}{1+b_{j}(1+d_{j})^{2}} \: .
    \end{equation}
  \end{enumerate}

\end{appendices}


\bibliography{DDT}

\end{document}